\documentclass[fleqn,usenatbib]{rasti}

\usepackage{newtxtext,newtxmath}
\usepackage[T1]{fontenc}

\DeclareRobustCommand{\VAN}[3]{#2}
\let\VANthebibliography\thebibliography
\def\thebibliography{\DeclareRobustCommand{\VAN}[3]{##3}\VANthebibliography}

\usepackage{graphicx}	% Including figure files
\graphicspath{{./Figures/}}
\usepackage{gensymb}	% degree symbol
\usepackage{amsmath}	% Advanced maths commands
\usepackage{color}

\newcommand*{\Dex}{\emph{DexRT}}
\newcommand*{\Lw}{\emph{Lightweaver}}
\newcommand*{\Pw}{\emph{Promweaver}}
\newcommand*{\Ha}{H$\alpha$}
\newcommand*{\Lyb}{Ly$\,\beta$}
\newcommand*{\CaK}{Ca$\,$\textsc{ii}$\,$K}
\newcommand*{\Caii}{Ca$\,$\textsc{ii}}
\usepackage{siunitx}
\DeclareSIUnit\gauss{G}

\title[Radiance Cascades for Non-LTE Radiative Transfer]{Radiance Cascades: A Novel High-Resolution
Formal Solution for Multidimensional
Non-LTE Radiative Transfer}

\author[Osborne \& Sannikov]{
C. M. J. Osborne,$^{1}$\thanks{E-mail: \href{mailto:Christopher.Osborne@glasgow.ac.uk}{Christopher.Osborne@glasgow.ac.uk}}
A. Sannikov$^{2}$
\\
$^{1}$SUPA School of Physics and Astronomy, University of Glasgow, G12 8QQ, UK\\
$^{2}$Grinding Gear Games, 6 Alderman Dr, Henderson, Auckland, New Zealand 0612\\
}

\date{Accepted XXX. Received YYY; in original form ZZZ}

\pubyear{\the\year{}}

\begin{document}
\label{firstpage}
\pagerange{\pageref{firstpage}--\pageref{lastpage}}
\maketitle

\begin{abstract}
Non-LTE radiative transfer is a key tool for modern astrophysics: it is the means by which many key synthetic observables are produced, thus connecting simulations and observations.
Radiative transfer models also inform our understanding of the primary formation layers and parameters of different spectral lines, and serve as the basis of inversion tools used to infer the structure of the solar atmosphere from observations.
The default approach for computing the radiation field in multidimensional solar radiative transfer models has long remained the same: a short characteristics, discrete ordinates method, formal solver.
In situations with complex atmospheric structure and multiple transitions between optically-thick and -thin regimes these solvers require prohibitively high angular resolution to correctly resolve the radiation field.
Here, we present the theory of radiance cascades, a technique designed to exploit structure inherent to the radiation field, allowing for efficient reuse of calculated samples, thus providing a very high-resolution result at a fraction of the computational cost of existing methods.
We additionally describe our implementation of this method in the \Dex{} code, and present initial results of the synthesis of a snapshot of a magnetohydrodynamic model of a solar prominence formed via levitation-condensation.
The approach presented here provides a credible route for routinely performing multidimensional radiative transfer calculations free from so-called ray effects, and scaling high-quality non-LTE models to next-generation high-performance computing systems with GPU accelerators.
\end{abstract}

% Include between one and six keywords.

%%%%%%%%%%%%%%%%%%%%%%%%%%%%%%%%%%%%%%%%%%%%%%%%%%

%%%%%%%%%%%%%%%%% BODY OF PAPER %%%%%%%%%%%%%%%%%%

\section{Introduction}

Radiative transfer calculations are a cornerstone of modern astronomy, modelling the propagation of light as it is affected by emission, absorption, and scattering processes, and eventually observable images and spectra.
This provides an essential connection between ever-increasingly advanced numerical simulations and observations, enabling scientific interpretation of the parameters of distant objects that cannot be measured \emph{in situ}.
Some regimes of emission and absorption allow for simplified solutions to this problem.
For example, in the optically thin regime, it is sufficient to consider only plasma emission properties along each line-of-sight, or in the regime of local thermodynamic equilibrium (LTE), the plasma emission and absorption properties for atomic\footnote{Herein we use ``atomic'' to refer to both neutral and ionised atoms.} spectral lines and continua can be computed from the local thermodynamic state, allowing for lines-of-sight to again be treated individually.
Many atomic spectral lines in bodies of astrophysical interest form outside of the LTE regime.
These are termed non-LTE.
We define three primary departures from LTE:
\begin{enumerate}
    \item Departures from a Boltzmann distribution for atomic excitation levels.
    \item Departures from a Saha distribution for atomic ionisation fractions.
    \item Departures from a Maxwellian velocity distribution.
\end{enumerate}
Most non-LTE methods are concerned with the (i) and (ii), however, \citet{paletou_full_2021} have recently started to develop approaches to additionally treat (iii).

Unlike in the LTE regime where the atomic ionisation and energy level distributions can be assumed to be dominated by collisional effects, non-LTE radiative transfer models the propagation of light through a domain where the emission and absorption coefficients are not known \emph{a priori}, and are affected by this self-same radiation field.
Absorption in one radiative transition at one location will therefore affect the atomic energy distribution at this location, and potentially influence the other transitions of this atom.
To this end, iterative approaches have been adopted for solving for the atomic ionisation and level populations (henceforth atomic populations or level populations) throughout the domain by first solving for the radiation field at a number of different wavelengths and computing the transition rates, before updating the atomic populations following these transition rates, and iterating this process until convergence.
For unpolarised radiation, the workhorse numerical methods of short characteristics \citep{kunasz_short_1988} for solving the radiative transfer equation for the radiation field, and a preconditioned treatment of the transition rates \citep{Rybicki1991,Rybicki1992}, have evolved little in the past 30 years.
A detailed overview of these techniques is provided in \citet{Hubeny2014}.

The core numerical problem of non-LTE radiative transfer in multiple dimensions is that of determining the radiation field throughout the domain, one key quantity of which is the first moment of the radiation field
\begin{equation}
    J_\nu(\vec{p}) = \frac{1}{4\pi}\oint_{\mathbb{S}^2} I_\nu(\vec{p}, \hat{\omega}) \mathop{d\hat{\omega}},
    \label{Eq:J}
\end{equation}
at frequency $\nu$ and location $\vec{p}$, and with $I_\nu$ the monochromatic specific intensity in direction $\hat{\omega}$.
This integral describes the gathering of light over all directions surrounding a point, with the potential for any other point in the domain to emit light that may be occluded en route to $\vec{p}$.
The angle-averaged $J_\nu(\vec{p})$, or indeed the full directional distribution of $I_\nu(\vec{p}, \hat{\omega})$, must be determined at every $\vec{p}$ to compute the atomic transition rates, and modification of emission and absorption coefficients at any point $\vec{p}$ may potentially affect the directional distribution of intensity at any other point.
Considering the steady-state solution to the radiative transfer equation, the general non-LTE problem is both non-linear and potentially extremely non-local.

This global and recursive problem does not occur uniquely in astrophysical radiative transfer but is also a core problem of energetic neutron transport in reactor design \citep[e.g.][]{boyd_openmoc_2014}, and the problem of global illumination in computer graphics.

In this work we present an approach for decomposing the radiation field, allowing for efficient reuse of calculated values, with a scaling that allows for the construction of exponentially more samples than those computed.
This is achieved by exploiting the structure of $I_\nu(\vec{p}, \hat{\omega})$, which is a five-dimensional function at each frequency $\nu$, but due to the propagation of light along straight paths (in the non-relativistic limit), it presents internal structure that can be exploited to produce a far more efficient representation termed \emph{radiance cascades}.

\subsection{Ray effects in multidimensional radiative transfer}\label{Sec:RayEffects}

\begin{figure*}
	\centering
    \includegraphics[width=\textwidth]{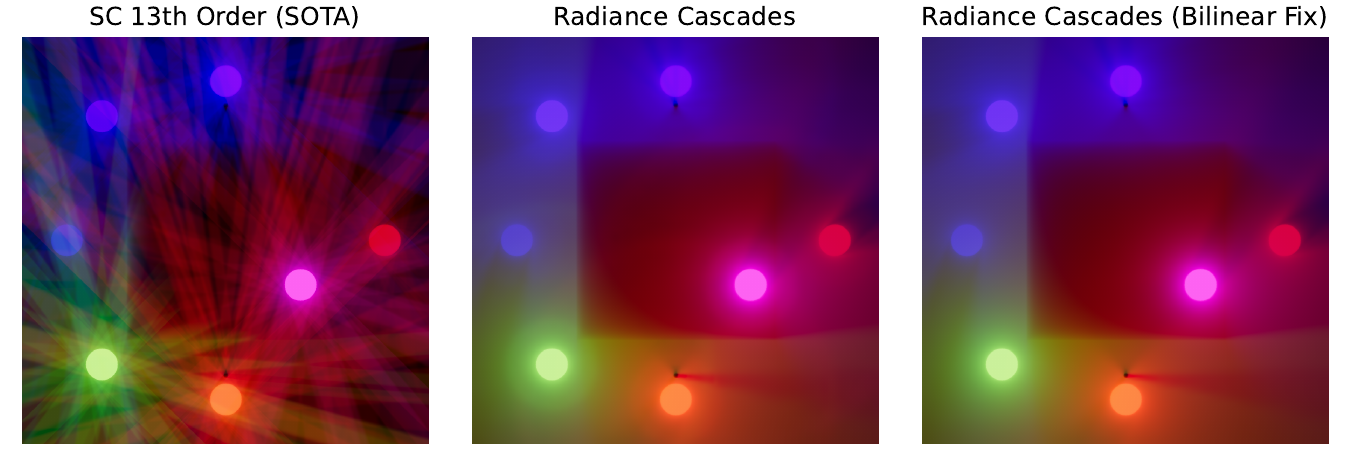}
    \caption{Comparison of the state-of-the-art short characteristics method (using a 10th order quadrature) with two different variants of the radiance cascades solution.
    For this scene, the ``Radiance Cascades (Bilinear Fix)`` panel may be taken as the ground truth.
    The scene is composed of opaque emissive circles of different intensities (at three different, independent wavelengths, represented by the red, green, blue triad), along with several absorbing features.
    For readability, the images are tonemapped from high-dynamic range into a visual proxy, enhancing the visibility of dim regions over a simple linear scaling.
    The large dark red box in the centre of the frame is weakly absorptive in red, and $100\times$ more absorptive in blue and green.
    The two small black circles near the blue and red circles at the top and bottom of the frame are strongly to all of the colour channels.
    The short characteristics solution presents very strong ray effects, demonstrating its unsuitability for problems that transition between finely-structured optically thick and optically thin regions.
    The basic radiance cascades method provides a very high-quality solution with minor ringing artefacts discussed in Section~\ref{Sec:RadianceCascadeArtefacts}, whilst the radiance cascades with bilinear fix solution effectively provides the expected ground truth solution.
    }
    \label{Fig:RcVsSc}
\end{figure*}

Methods for evaluating \eqref{Eq:J} can be categorised into two types: (i) stochastic (Monte Carlo), and (ii) deterministic techniques.
Stochastic path tracing methods are a cornerstone of modern computer graphics \citep{pharr_physically_2023}, but suffer from issues with convergence at high optical depths present in many astrophysical models \citep{camps_failure_2018}.
Additionally, whilst these methods rapidly provide an approximate solution in many settings, their stochastic approach yields a reduction in noise proportional to the square-root of the number of samples taken.

% NOTE: SN and PN are equivalent in 1D with SN using a Gauss-Legendre quadrature.
Deterministic techniques are typically split into $P_N$ and $S_N$ methods as different approaches for computing the angular integral \citep{marguet_computational_2017}.
The former expands the intensity distribution into spherical harmonics, truncated at an order $N$.
Due to the coupling between the different angular terms this introduces, the order is typically kept low, and used to model large sources with relatively isotropic intensity distributions.
The $S_N$, or discrete ordinates method, is the primary technique employed for non-LTE radiative transfer.
It consists of computing the angular integral via a set of $N$ discrete samples at weighted angles: the most common choice being the sets described by \citet{carlson_numerical_1963,lathrop_discrete_1964}.
This method is known to fail, producing so-called \emph{ray effects} clearly visible in media without scattering \citep{lathrop_ray_1968}, due to the purely discrete angular samples of the intensity distribution taken, but these effects are dramatically reduced in diffusive media.
The choice of angular quadrature is therefore of the utmost importance.
\citet{Stepan2020} commented on the failure of the Carlson quadrature set to correctly integrate weakly polarised radiation fields, whilst \citet{jenkins_bright_2024} were forced to adopt a uniform solid angle quadrature with 96 rays/octant based on HEALPix \citep{gorski_healpix_2005} to correctly integrate the radiation field between the solar surface and a prominence model due to the high albedo of the overlying structure.
This latter treatment leads to an extension of the $S_N$ technique by considering each sample as the approximation of the average over a cone with associated solid-angle rather than an infinitesimal sample.

In a discrete ordinates method, the obvious method of evaluating \eqref{Eq:J} is to trace $N$ rays from each $\vec{p}$ along the directions of the quadrature, solving the radiative transfer equation along each, until the edges of the domain.
This is known as the method of long characteristics \citep[e.g.][]{jones_formation_1973,jones_formation_1973-1}, and is extremely computationally expensive.
Instead, the short characteristics approach of \citet{kunasz_short_1988} is commonly adopted, where the radiation field is swept in the direction of transport across the grid, with each $\vec{p}$ finding its upstream value along a ray by interpolation.
The choices of interpolation both along the ray and across the grid are key to avoiding numerical instability and over-diffusion \citep{auer_two-dimensional_1994}.

Figure~\ref{Fig:RcVsSc} shows a comparison between a discrete ordinates method using the 10 ray/octant (accurate to 13th order spherical harmonics) unpolarised quadrature of \citet{Stepan2020} and our new radiance cascades method, on a test problem involving optically thick sources, occluders, and absorption.
The ``Radiance Cascades (Bilinear Fix)'' panel may be taken as the ground truth for this model.
The test problem is computed in three different wavelengths, then presented as the red, green, and blue colour channels.
It is composed of a number of opaque emissive circles, two small opaque absorbing circles (near the red and blue circles at the bottom and top of the frame, respectively), and a central square absorbing region.
This region is weakly absorbing in red, and $100\times$ more absorptive in green and blue.
The remaining space is pure vacuum with no emissive or absorbing effects.
The short characteristics method (computed with a BESSER formal solver and BESSER spatial interpolation \citep{Stepan2013} using \Lw{} \citep{Osborne2021}) exhibits very strong ray effects that are not present in our radiance cascades solutions.
It is theoretically possible to increase the angular quadrature sufficiently to create a smooth solution as in the radiance cascades cases, but the computational cost scales linearly with the number of angular samples employed.
Correctly resolving distant sources will require extremely high angular resolution due to the small effective size of these sources, however this angular resolution is not needed for nearby sources, where instead high spatial resolution is needed.
It is this tradeoff that our radiance cascades method exploits to achieve the necessary angular resolution with dramatically higher efficiency.

\subsection{The \Dex{} Code}

The \Dex{} code described herein is an open-source implementation of the radiance cascades method applied to non-LTE radiative transfer in atomic spectral lines.
It is available on GitHub\footnote{\url{https://github.com/Goobley/DexRT/}} with archival on Zenodo\footnote{\url{https://zenodo.org/doi/10.5281/zenodo.13376008}}, under the Apache-2.0 license\footnote{\url{https://opensource.org/license/apache-2-0}}.

\section{Radiance Cascades: Motivation \& Construction}\label{Sec:Method}

\emph{The core radiance cascades method was developed by A. Sannikov (in prep)\footnote{A preprint is available here: \url{https://github.com/Raikiri/RadianceCascadesPaper}.} primarily for use in realtime computer graphics such as video games, with the application to non-LTE radiative transfer by C. M. J. Osborne}

In Section~\ref{Sec:PenumbraCriterion} we motivate the construction of this technique through the trade-off between necessary angular- and spatial-resolution when sampling the radiation field.
Then, in Sections~\ref{Sec:RadianceIntervals}--\ref{Sec:InterpolatedRayAnalysis} we describe how these properties are exploited by the radiance cascades method.
Finally, in Section~\ref{Sec:RadianceCascadeArtefacts} we discuss the origin of some artefacts that result from this method, along with potential mitigation techniques.

\subsection{Motivation - Penumbra Criterion}\label{Sec:PenumbraCriterion}

\begin{figure}
	\centering
    \includegraphics[width=\columnwidth]{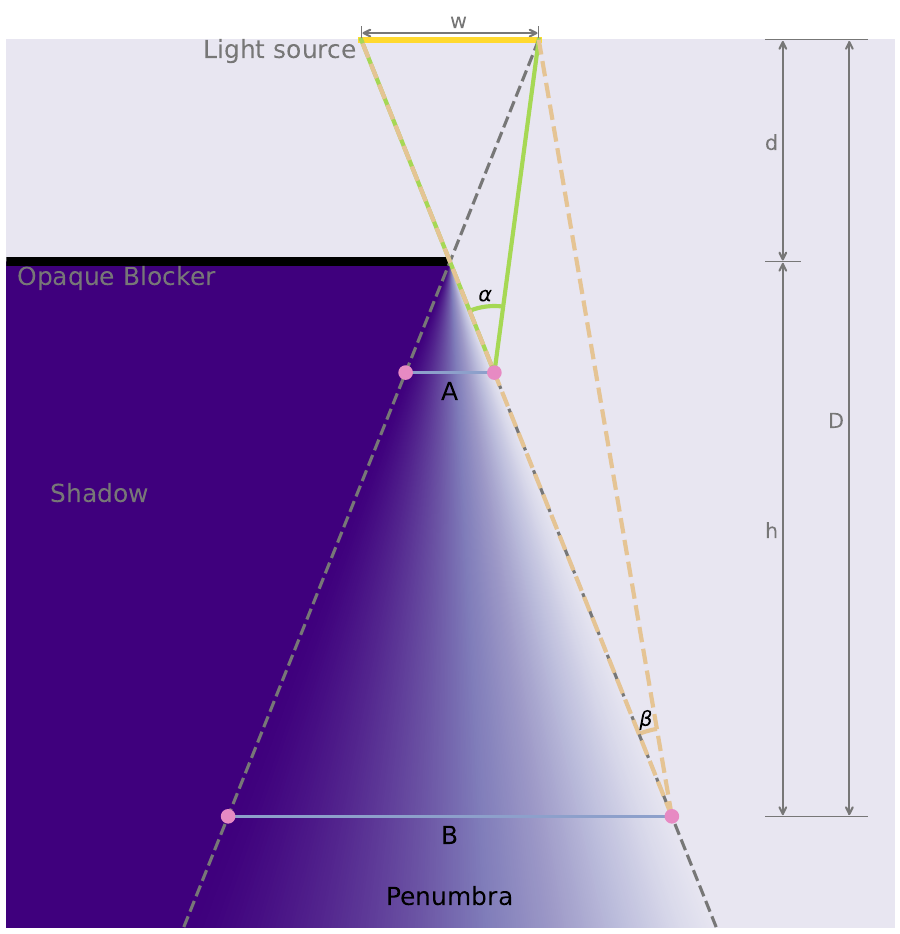}
    \caption{Illustration of the penumbra criterion motivating the radiance cascades method.
    %  The labelled $h$ and $D$ distances are shown for the probes on segment $\mathrm{B}$.
     Considering a linear light source illuminating a two-dimensional plane with a blocker casting a shadow we observe that points close to the blocker (on the segment $\mathrm{A}$) require high spatial- but low angular-resolution to resolve the illumination of the penumbra, whereas those far from the blocker (on the segment $\mathrm{B}$) require lower spatial- but higher angular-resolution.}
    \label{Fig:PenumbraCriterion}
\end{figure}

To motivate the construction of our new formal solution, let us consider the properties of a simple two-dimensional flatland problem, shown in Figure~\ref{Fig:PenumbraCriterion}.
The labelled distances on this figure are shown for the points on segment $\mathrm{B}$ but generalise to $\mathrm{A}$ too.
A simple emissive line segment illuminates a region of free space with a completely opaque blocker.
Naturally, this blocker casts a shadow.
Due to the non-negligible angular size of the light source there is a region of \emph{penumbra}, or partial shadow, where the light received is proportional to the angle (in 2D, solid angle in 3D) subtended by the light source that is visible from the point.

Let us consider resolving the penumbra at two different distances from the blocker: these near and far setups are shown by the points on the line segments $\mathrm{A}$ and $\mathrm{B}$, respectively.
On both of these line segments, two pink points are indicated: these are on the borders of the penumbra and represent the closest points along these lines that are fully shadowed and fully illuminated.
Between these points the brightness of the penumbra varies near-linearly as the illumination is simply determined by the angular size of the fraction of the light source.
When comparing these scenarios, we can remark that close to the blocker the radiation field changes rapidly in space and points must be placed close together, whereas further from the blocker, the radiation field varies smoothly on a much larger spatial scale ($\mathrm{A} < \mathrm{B}$).
Additionally, close to the blocker, the light source subtends a much larger angle than far from the blocker ($\alpha > \beta$).

Considering the geometry of the situation, for a light source of length $w$ at perpendicular distance $d$ from the blocker a penumbra of angular size

\begin{equation}
    \gamma = 2\arctan{\frac{w}{2d}}
\end{equation}

is cast.
The penumbra at perpendicular distance $h$ from the blocker then has linear size
\begin{equation}
    H(h) = 2 \arctan{\left(\frac{w}{2d}\right)} h = \gamma h.
\end{equation}
This provides a condition on the spatial sampling necessary to resolve the penumbra
\begin{equation}
    \Delta_s < H(h).
\end{equation}
There is also a necessary condition for our method to angularly resolve the light source, which at distance $D$ subtends an angular size of
\begin{equation}
    \epsilon(D) = 2 \arctan{\left( \frac{w}{2D} \right)},
\end{equation}
giving a necessary angular resolution of
\begin{equation}
    \Delta_\omega < \epsilon(D).
\end{equation}

To provide a simple unification of these requirements, consider the case where $d \ll D$ and $\gamma$ is small, giving
\begin{equation}
\begin{cases}
    h \approx D, \\
    \Delta_s \lesssim \gamma D, \\
    \Delta_\omega \lesssim w / D.
\end{cases}
\end{equation}

These observations can then be distilled into the penumbra criterion:
\begin{itemize}
    \item Near-field radiance contributions vary with high spatial frequency and low angular frequency.
    \item Far-field radiance contributions vary with low spatial frequency and high angular frequency.
\end{itemize}
Mathematically,
\begin{equation}
    \begin{cases}
        \Delta_s < F(D) \propto D, \\
        \Delta_\omega < G(1/D) \propto 1 / D,
    \end{cases}
\end{equation}
where $F(D)$ and $G(1/D)$ are arbitrary linear functions.
In the case where $d \approx D$, and the angle of the penumbra is not small, an equivalent penumbra criterion may still be constructed, but requires that $\Delta_\omega$ scale superlinearly with $1 / D$.

We hypothesise that the penumbra criterion holds generally.
Consider the same situation but where the light source is located far from the occluder: the shadow will be sharper, i.e. the penumbra will be narrower as the light source subtends a smaller angle.
In this situation points both close and far from the blocker require high angular-resolution to resolve the distant light-source, while the points close to the blocker require high spatial-, but low angular-resolution to resolve the blocker and thus the penumbra.
Radiance cascades are designed to capture the variation of the radiance field at different distances from each point within the constraints of the penumbra criterion, considering arbitrary emitting and occluding features.

\subsection{Radiance Intervals}\label{Sec:RadianceIntervals}

\begin{figure*}
	\centering
    \includegraphics[width=\textwidth]{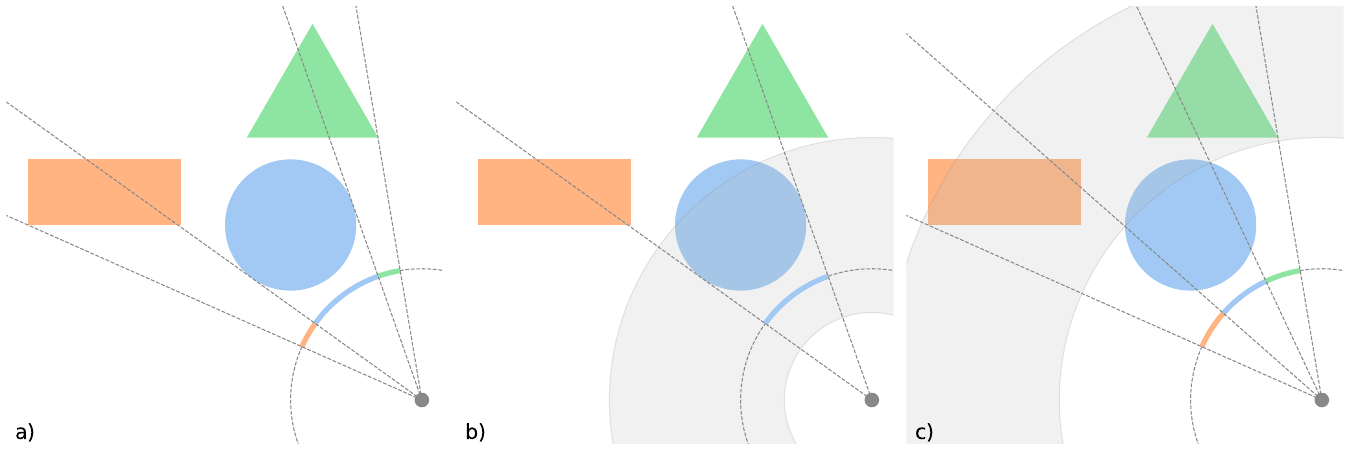}
    \caption{Comparison of long characteristics style ray-casting from a point against two different radiance intervals over the same field. The three coloured primitives can be considered opaque emissive sources. In a) we show the a conventional ray-casting approach, whilst b) and c) show the radiance found inside annuli described by closer and further radiance intervals (shown in grey). Note that in b) only the blue circle is found by these samples, whilst in c) obscured components of the orange rectangle and green triangle are sampled.}
    \label{Fig:RadianceInterval}
\end{figure*}

The primary building block of radiance cascades is the radiance interval.
For an arbitrary volume of emitting and absorbing media, considering the specific intensity at a single wavelength, we denote $\mathcal{R}_{a,b}(\vec{p}, \hat{\omega})$ the two-vector of monochromatic specific intensity $I(\vec{p}, \hat{\omega})$ and monochromatic optical depth $\tau(\vec{p}, \hat{\omega})$ accumulated along the path $\vec{P}(t) = \vec{p} + t\hat{\omega}$ for position $\vec{p}$, direction $\hat{\omega}$, and $t\in[a, b]$, i.e.
\begin{equation}
    \mathcal{R}_{a,b}(\vec{p}, \hat{\omega}) = [I_{b\rightarrow a}(\vec{p}, \hat{\omega}), \tau_{b\rightarrow a}(\vec{p}, \hat{\omega})],
    \label{Eq:Rab}
\end{equation}
where quantities with the subscript ${b\rightarrow a}$ can be thought of as ``accumulated along the path from $b$ to $a$''.
$\mathcal{R}_{a,b}(\vec{p}, \hat{\omega})$ can be integrated by the application of any traditional formal solver along this path.
A radiance interval therefore encodes a specific intensity that may illuminate $\vec{p}$, but also the effectiveness of this path to occlude radiation from further upwind.
A collection of radiance intervals with given angular sampling can then encode the potential illumination received by a point from a shell of distance $[a, b]$.
Note that the definition of each radiance interval is not unique, as
\begin{equation}
    \mathcal{R}_{a+x,b+x}(\vec{p}, \hat{\omega}) = \mathcal{R}_{a, b}(\vec{p} + x\hat{\omega}, \hat{\omega}).
    \label{Eq:RabMerge}
\end{equation}

A comparison of the conventional long characteristics ray-casting approach and two annuli of radiance intervals around a point sampling a scene of opaque primitives (circle, rectangle, and triangle) is shown in Figure~\ref{Fig:RadianceInterval}.
In a discretised solution, the long characteristics ray-casting approach attempts to encode the radiation field from all distances using a single angular quadrature, leading to unnecessary oversampling for nearby sources, and potential undersampling of distant sources.
The two different shells described by radiance intervals only consider interactions with primitives inside their range and a different optimised quadrature can therefore be used for each of these to satisfy the penumbra criterion for each radiance interval.

\subsubsection{Radiance Interval Merging}\label{Sec:Merging}

Contiguous radiance intervals can be merged via the merging operator $\mathcal{M}$ such that
\begin{equation}
    \begin{split}
    \mathcal{R}_{a,c}(\vec{p}, \hat{\omega}) &= \mathcal{M}\left(\mathcal{R}_{a,b}(\vec{p}, \hat{\omega}), \mathcal{R}_{b,c}(\vec{p}, \hat{\omega})\right)\\
    &\begin{split}
           \:= [&I_{c\rightarrow b}(\vec{p}, \hat{\omega}) + \exp{(-\tau_{c\rightarrow b}(\vec{p}, \hat{\omega}))}I_{b\rightarrow a}(\vec{p},\hat{\omega}),\\
           &\tau_{c\rightarrow b}(\vec{p}, \hat{\omega}) + \tau_{b\rightarrow a}(\vec{p}, \hat{\omega})].
        \end{split}
    \end{split}
\end{equation}
This procedure for merging intervals arises naturally from the radiative transfer equation.

\subsection{Radiance Cascades}\label{Sec:RadianceCascades}

The question now arises of how to use radiance intervals to exploit the properties of the penumbra criterion discussed in Section~\ref{Sec:PenumbraCriterion} and thus efficiently encode the radiation field in arbitrarily complex models with many interfaces between optically-thin and -thick structures.

From the penumbra criterion, to correctly resolve the radiation coming from a shell $[a,b]$ where $b > a$, we require a spatial resolution $\Delta_s = \min(a, b) = a$, and an angular resolution of $\Delta_\omega = \min(1/a, 1/b) = 1/b$.
We then employ a set of radiance intervals to determine the radiation at a point $\vec{p}$ from each surrounding shell.
The set of radiance intervals representing a particular shell for each point in the model is termed a \emph{radiance cascade}, i.e. a radiance cascade encodes $\mathcal{R}_{a,b}(\vec{p},\hat{\omega})$ for a given $[a, b]$, over all $\vec{p}$ in the model and $\hat{\omega} \in \mathbb{S}^2$.
A sufficient set of radiance cascades can be constructed by ensuring that the penumbra criterion remains valid for each cascade, and a natural definition for this is to scale each subsequent cascade with exponential doubling, giving a definition for cascade $i$ covering a shell $[t_i, t_{i+1}]$ from each probe centre
\begin{equation}
\begin{cases}
    \Delta_s \propto 2^{i}, \\
    \Delta_\omega \propto \frac{1}{2^{\alpha i}},\\
    t_i \propto 2^{\alpha i},
\end{cases}
\label{Eq:RcScalingLaw}
\end{equation}
where $\alpha \geq 1$ is known as the cascade branching factor.
Different branching factors can naturally be applied to all three terms whilst satisfying the penumbra criterion, but in practice we have found good efficiency by applying it to the angular sampling and radiance interval lengths, whilst leaving the sample spacing increasing by successive doubling.
Note that the values of these parameters on the shortest range cascade, from which all others are scaled, are free parameters and must be set so that the penumbra criterion is met, as such we require
\begin{equation}
    \begin{cases}
        t_0 = 0, \\
        \lim_{i\rightarrow\infty} t_i = \infty.
    \end{cases}
\end{equation}
In practice satisfactory expressions for $\Delta_s$ and $\Delta_\omega$ on the zeroth cascade can be quickly found by visual inspection of the solution, taking into account the spatial scales of the problem.

\begin{figure}
	\centering
    \includegraphics[width=\columnwidth]{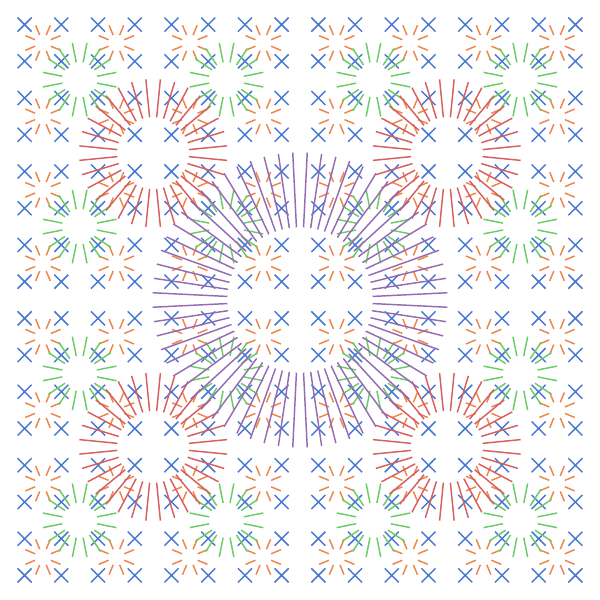}
    \caption{A set of five radiance cascades with following the scaling laws of \eqref{Eq:RcScalingLaw} with branching factor $\alpha=1$. These cascades (ordered blue, orange, green, red, purple) capture progressively higher angular-resolution representations of increasingly distant radiation field components at decreasing spatial resolution, thus exploiting the properties of the penumbra criterion.}
    \label{Fig:RcGrid}
\end{figure}

A simple approach that satisfies these discretisation criteria is to encode radiance intervals in probes placed in a uniform rectilinear grid, with regularly spaced directional intervals.
A set of five radiance cascades following this structure (with $\alpha=1$) is shown in Figure~\ref{Fig:RcGrid}.
The shortest range cascade is shown in blue, starting at each probe location: the points where $J$ is to be determined, and utilises $4$ angular samples.
The next cascade, with twice the spacing, is shown in orange, and starts at a distance of $t_0$ away from each probe location and contains $2\alpha \times 4=8$ angular samples per probe.
To correctly resolve boundary conditions the rays in the upper-most cascade typically extend until they leave the simulation domain.
This does not violate the penumbra criterion provided the cascade naturally has enough angular resolution to encode the far-field radiation field of the boundary.
If this is not the case, additional cascades are added until this condition is met.

\subsubsection{Interpolated Ray Construction}\label{Sec:RayInterpolation}

From the previous section, our set of radiance cascades, as illustrated in Figure~\ref{Fig:RcGrid} encodes the radiation field throughout the model.
This does not, however, give a value for the angle-averaged radiation field $J$ at every point in the model.
The determination of $J$ at each point is computed by interpolation and merging of the cascades.
In Section~\ref{Sec:PenumbraCriterion}, we remarked that the intensity between the fully obscured and fully illuminated probes on segments $A$ and $B$ of Figure~\ref{Fig:PenumbraCriterion} varied linearly, and indeed, the information encoded in the radiance cascades can be linearly interpolated in both space and direction.
This is due to the spatial and angular resolutions meeting their specific Nyquist criteria: higher frequency information is removed due to the design of the cascades.
For example, a probe in cascade 0 does not need to and, indeed, cannot encode a field with finely-structured angular variation as it only includes contributions from near-field radiance.
On the other hand, probes in an upper cascade cannot encode fine spatial variations, as they only encode contributions from far-field radiance which varies slowly in space.

Probes from higher cascades are interpolated onto lower cascades by $n$-linear interpolation.
In the case of our 2D example, this is bilinear interpolation from four probes of cascade $i+1$ onto one probe of cascade $i$.
As the relative positions of the probes are well known, the weights for this bilinear interpolation can be computed very efficiently.

\begin{figure}
	\centering
    \includegraphics[width=\columnwidth]{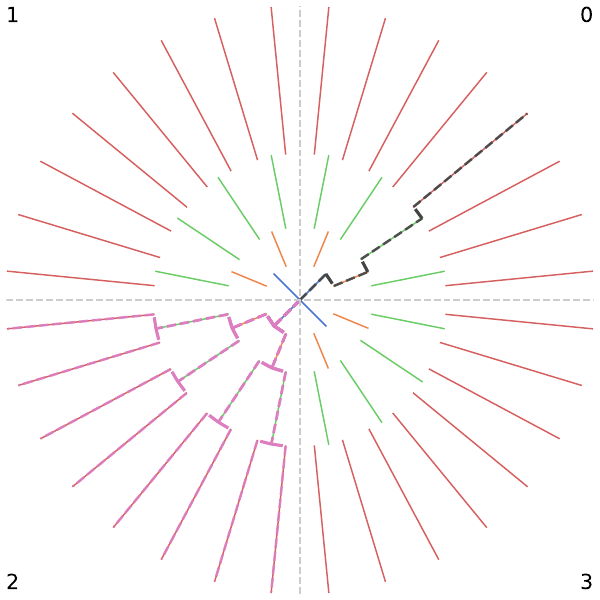}
    \caption{Five radiance cascades, interpolated on one probe of cascade 0. The colour of each cascade is the same as Figure.~\ref{Fig:RcGrid}. In quadrant 0 a single ray constructed by cascade interpolation and merging is highlighted in dashed black. The constructed rays that make up quadrant 2 are highlighted in pink.}
    \label{Fig:InterpCascadeHighlight}
\end{figure}

An illustration of the five cascades from Figure~\ref{Fig:RcGrid} interpolated onto a single probe of cascade 0 is shown in Figure~\ref{Fig:InterpCascadeHighlight} with one effective ray of quadrant 0 highlighted in dashed black.
The highlighted interpolated radiance intervals are then merged following \eqref{Eq:RabMerge}.
We note that following interpolation these radiance intervals share a common origin $\vec{p}$ but are not contiguous in angle $\hat{\omega}$.
In practice this is not an issue for determining $J$, because, as discussed in Section~\ref{Sec:RayEffects}, each radiance interval may serve as an approximation of the radiance field in a cone, and each radiance interval of cascade $i$ has $2\alpha$ ``child'' rays in this same cone in cascade $i+1$.
The recursive merging of all ``child'' rays of the single ray of cascade 0 in quadrant 2 is highlighted in dashed pink.
When considering a cone such as this, the jaggedness of the single highlighted black ray in quadrant 0 is no longer so apparent.
As our primary quantity of interest is the angle-averaged $J$, we do not anticipate this introducing meaningful error so long as the tunable cascade 0 parameters are correctly configured (see Section~\ref{Sec:RadianceCascades}).

If one instead wishes to determine the specific intensity along a particular infinitesimal angular ray, then the upper cascades can be interpolated in angle prior to merging.

Implementations of the radiance cascades method can perform this interpolation and merging step iteratively: after calculation, each interval of cascade $i$ merges with the $2\alpha$ ``child'' intervals of cascade $i+1$.
Cascade $i-1$ is then computed and merged with cascade $i$.

\subsection{Interpolated Ray Analysis}\label{Sec:InterpolatedRayAnalysis}

For the discussion of interpolated rays we remain in the two-dimensional flatland setup, but stress that this analysis also applies to a general three-dimensional setup.

Considering flatland cascades 0 and 1, with $\alpha=1$.
In cascade 0 we have $P_0$ probes, with $W_0$ angular samples taken at each probe, giving the number of rays to be computed in cascade 0, $N_{C_0} = P_0 W_0$.
For cascade 1, the spatial resolution is halved on each axis, i.e. $P_1 = P_0 / 4$ whilst the angular resolution is doubled: $W_1 = 2 W_0$, thus $N_{C_1} = P_1 W_1 = N_{C_0} / 2$.
Now, by interpolation, we can construct $W_1$ rays of cascade 1 at each probe in cascade 0, giving an effective number of rays of length $t_1$ constructed $N^\prime_{C_1} = P_0 W_1$.

We note that we have computed $\sum_i N_{C_i} = N_{C_0} + N_{C_1} = 3 N_{C_0} / 2$ rays, but by interpolation and merging have constructed $N^\prime_{C_1} = 2 N_{C_0}$ rays.
This behaviour continues as we add additional cascades: each cascade requires half of the rays of the previous, but doubles the effective angular sampling of the outermost radiance intervals by a factor of two, whilst interpolating this onto the spatial sampling of cascade 0.
In general, for $I$ cascades this gives

\begin{equation}
\begin{cases}
\begin{split}
\mathrm{Rays\,computed}&= \sum_i^I N_{C_i} \\
    &= N_{C_0} + N_{C_1} + \ldots + N_{C_i}\\
    &= \left(1 + \cfrac{1}{2} + \ldots + \cfrac{1}{2^i}\right) N_{C_0} \\
    &< 2 N_{C_0},
\end{split}\\
\begin{split}
\mathrm{Rays\,constructed}&=\,N^\prime_I\\
    &= 2^I N_{C_0}.
\end{split}
\end{cases}
\end{equation}

This asymptotic scaling breaks down before allowing the construction of infinite rays at finite cost, as for a given discretised model there are limits on the angular sampling needed to resolve the far-field radiance.
However, whilst the penumbra criterion holds, the scaling of constructed rays also holds, allowing for the integration of the radiation field at high spatial- and angular-resolution far cheaper than could be obtained with any traditional ray-casting approach where a doubling of angular resolution requires a doubling of the number of rays.

We note that the number of rays computed does not necessarily linearly scale with the computing time necessary for a solution: the radiance intervals in upper cascades are longer than those in lower cascades, and a full solution to the radiative transfer equation along their length can require more computational effort than for a short ray.
Nonetheless, this scaling remains very powerful, and multiple approaches can be taken to accelerate the longer rays.

This scaling law is also affected by the branching factor $\alpha$.
For example, in two-dimensions with $\alpha=2$ each cascade contains the same number of rays, i.e. $N_{C_i} = N_{C_0}$, so the total number of rays computed becomes $IN_{C_0}$, whilst the number constructed by interpolation becomes $2^{\alpha I}N_{C_0}$.
As such, the exponential scaling remains.
In all examples presented in this work we employ $\alpha=2$.

\subsection{Artefacts}\label{Sec:RadianceCascadeArtefacts}

\begin{figure*}
	\centering
    \includegraphics[width=\textwidth]{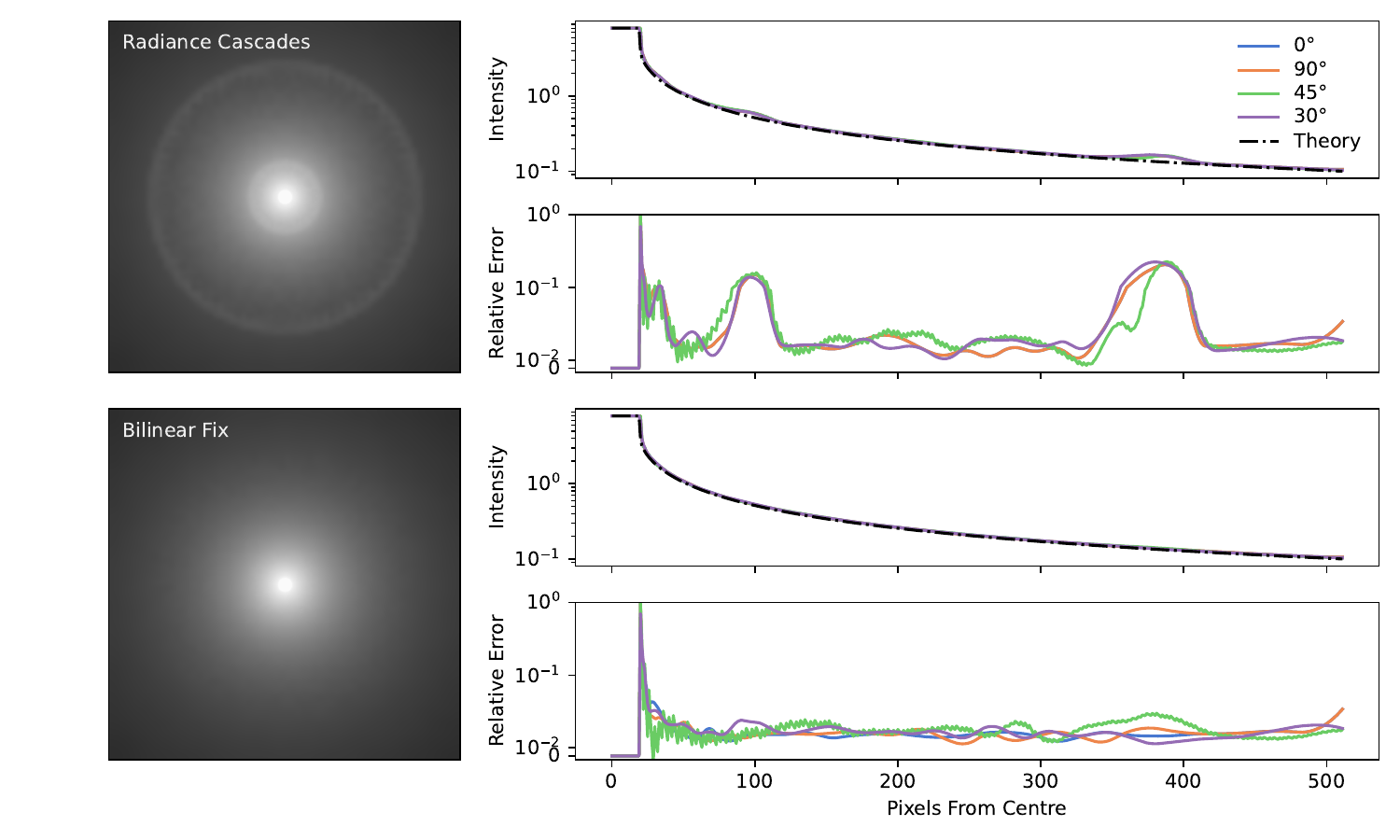}
    \caption{Comparison of falloff around an opaque circular source using basic radiance cascades, and those with the bilinear fix. On the left-hand side, a tonemapped representation of the radiation field is shown. The upper panel of each pair on the right-hand side is a comparison of the intensity falloff against a theoretical model shown for cuts at different angles starting from the centre of the source, whilst the lower panel of each pair is the relative error between the solution along these different cuts and the expected falloff. We note that this is the worst-case scenario for ringing from the basic radiance cascades method as it shows an extremely opaque source embedded in a completely transparent medium.}
    \label{Fig:CircleTest}
\end{figure*}

\begin{figure}
	\centering
    \includegraphics[width=\columnwidth]{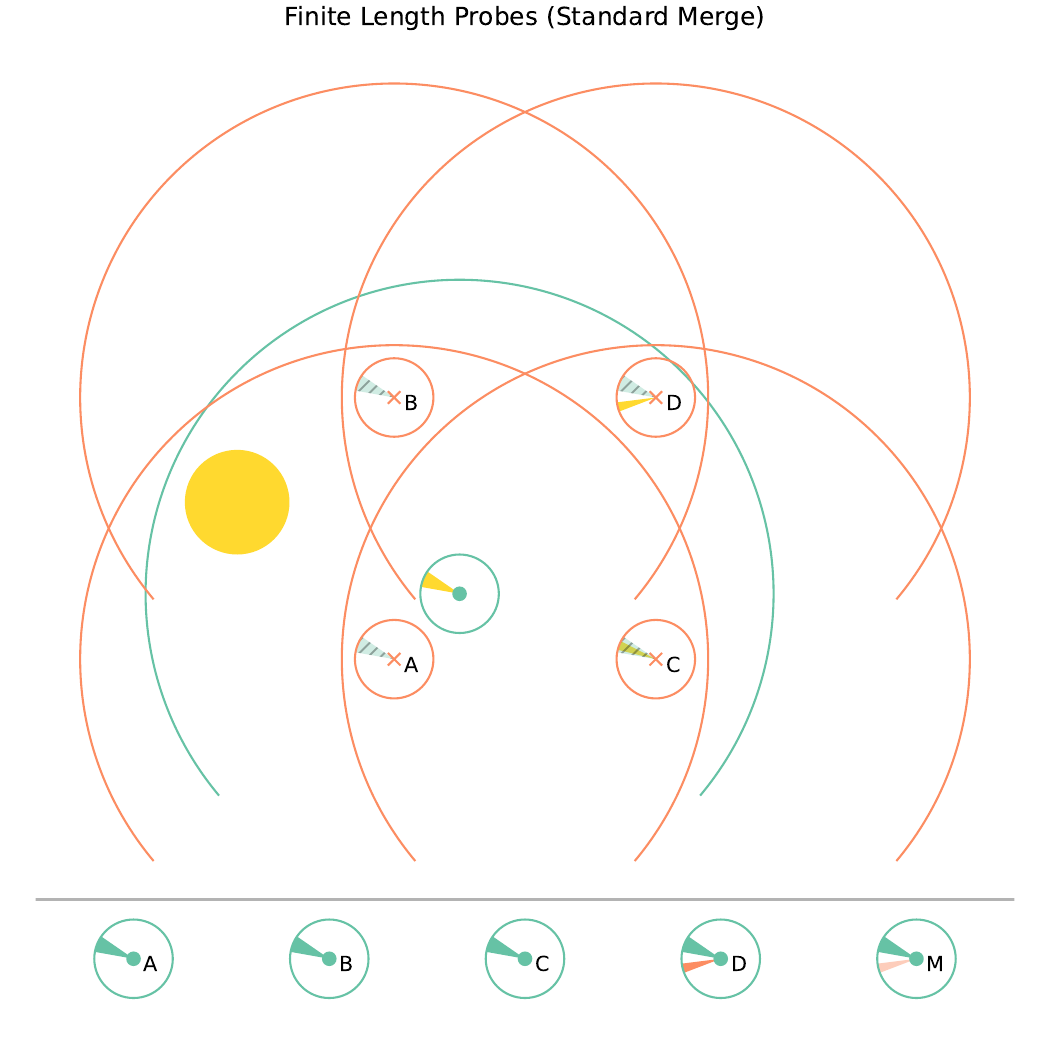}
    \caption{Illustration of the scenario in the basic radiance cascades method that leads to the ringing artefact. At the top of the graphic a probe of cascade $i$ is shown in green, and the four bilinearly sampled probes of cascade $i+1$ are shown in orange. The rings surrounding the probes indicate the end of the radiance interval associated with the probe of cascade $i$, and the start of the radiance interval associated with cascade $i+1$. Both the probe of cascade $i$ and probes C and D of cascade $i+1$ see the yellow light source, as indicated by the yellow cone. The opacity in the probe of cascade $i$ prevents merging the hatched area. The lower row provides an illustration of the result of merging the probe of cascade $i$ with each probe of cascade $i+1$, and the bilinear interpolation of these indicated by M. Note the additional source that isn't occluded in D and M, this is the origin of the ringing artefact.}
    \label{Fig:ParallaxStandard}
\end{figure}

For correctly chosen parameters, as shown in Figure~\ref{Fig:RcVsSc}, the radiance cascades solution eliminates the ray effects associated with classical discrete ordinates methods by dramatically increasing the angular resolution of the far-field.
This method presents a different form of artefact which is visible in the middle panel of Figure~\ref{Fig:RcVsSc}: a slight ringing that is visible around bright optically thick sources, embedded in optically thin layers.
A simple test case with a single light source illustrating this ringing is shown in Figure~\ref{Fig:CircleTest}.
Due to the tonemapping of this image (increasing the effective brightness of dim regions), the rings are very easily visible.
Consulting the relative error plot (second from the top on the right-hand column), we see that the error extends up to slightly over 10\%.
This ringing occurs where the different cascades meet, and this test represents the worst-case scenario of a small, very high opacity source (100 mean free paths per pixel, in a 1024$\times$1024 image) embedded in a completely transparent medium.
In more realistic cases for radiative transfer of spectral lines in sufficiently resolved models, the errors associated with ringing are typically less than a few per cent -- comparable or lower than the error associated with the interpolation schemes used in multidimensional short characteristics.
We note that the radiance cascades solution always lies above the theoretically predicted solution (see the relative error plots in the right-hand column), this is due to the discretisation of the circle into pixels used in the input to the model: any pixel partially through which the boundary of the circle passes is fully included as part of the opaque source.
This also explains the high error immediately next to the source, which falls off very rapidly over the first few pixels.
The error rising sharply (but remaining low) in the $[500,512]$ region for the $0\degree$ and $90\degree$ lines is due to the lack of full bilinear coverage of the upper cascades close to the edge of the grid.
Outside the outermost row or column of probes in a cascade, we clamp the lookup and effectively perform linear interpolation along the axis parallel to the edge, but lose the falloff due to there being no further probes.

The ringing artefact can be addressed by the use of the so-called bilinear fix (shown in the lower panels of Figure~\ref{Fig:CircleTest}) whereby each radiance interval in cascade $i$ is traced from its usual starting position to the starting position of the associated child cone on each of the four probes of cascade $i+1$ that are normally bilinearly interpolated and then merged with this interval.
Instead, following the bilinear fix, the four distinct radiance intervals are each merged with their associated samples from cascade $i+1$, and then averaged using the usual bilinear weights.
This leads to a $4\times$ increase in the number of rays that need to be traced on every cascade except cascade $I$, however, in addition to the ringing, the bilinear fix also addresses potential issues related to light leaking through an occluder that is too small.
That said, light leaking artefacts are typically an indication that the penumbra criterion is violated for some property of the model (e.g. the spacing of cascade 0 probes is too large).
In practice, this leaking is not an issue in the models considered here, as we apply the method to the output of magnetohydrodynamic (MHD) models, with a cascade 0 probe in every cell where $J$ is to be found, and such small features (one grid cell or smaller) should not occur if the underlying model correctly resolves the thermodynamic structure.

The ringing in the method originates from a situation where both a probe in cascade $i$, and some of the bilinear probes sampled in cascade $i+1$ sample the same source.
This is illustrated in Figure~\ref{Fig:ParallaxStandard}: the green probe shows the probe of cascade $i$ and the four orange probes are in cascade $i+1$.
The result of the merge with each of the four probes of $i+1$ is shown in the lower row, along with the final interpolated result (labelled M) where the opacity of the samples is set by their bilinear weights.
Probes A and B do not sample the source, as they are closer than the probe of cascade $i+1$ and their radiance intervals start at the same distance as the end of the interval associated with the probe of cascade $i$.
Due to their spatial offset however, probes C and D \emph{do} sample the source.
The directional alignment between the probe of cascade $i$ and C is good, and so this contribution is occluded during the merging process.
This is not the case for probe D, which accumulates the contribution from the light source at a different angle to the occlusion from the probe of cascade $i$.
As such, the result of merging (shown as M) the probe of cascade $i$ with the four bilinearly interpolated probes of cascade $i+1$ is contains two sources corresponding to this light source: the green contribution from the probe of cascade $i$ which contains the expected irradiance contribution from a probe at this distance, in addition to an extra contribution from probe D.
The ringing artefact is therefore due to parallax between the the different probes of cascade $i+1$ and cascade $i$, leading to non-conservation of energy at locations where cascades overlap a light source.

\begin{figure}
	\centering
    \includegraphics[width=\columnwidth]{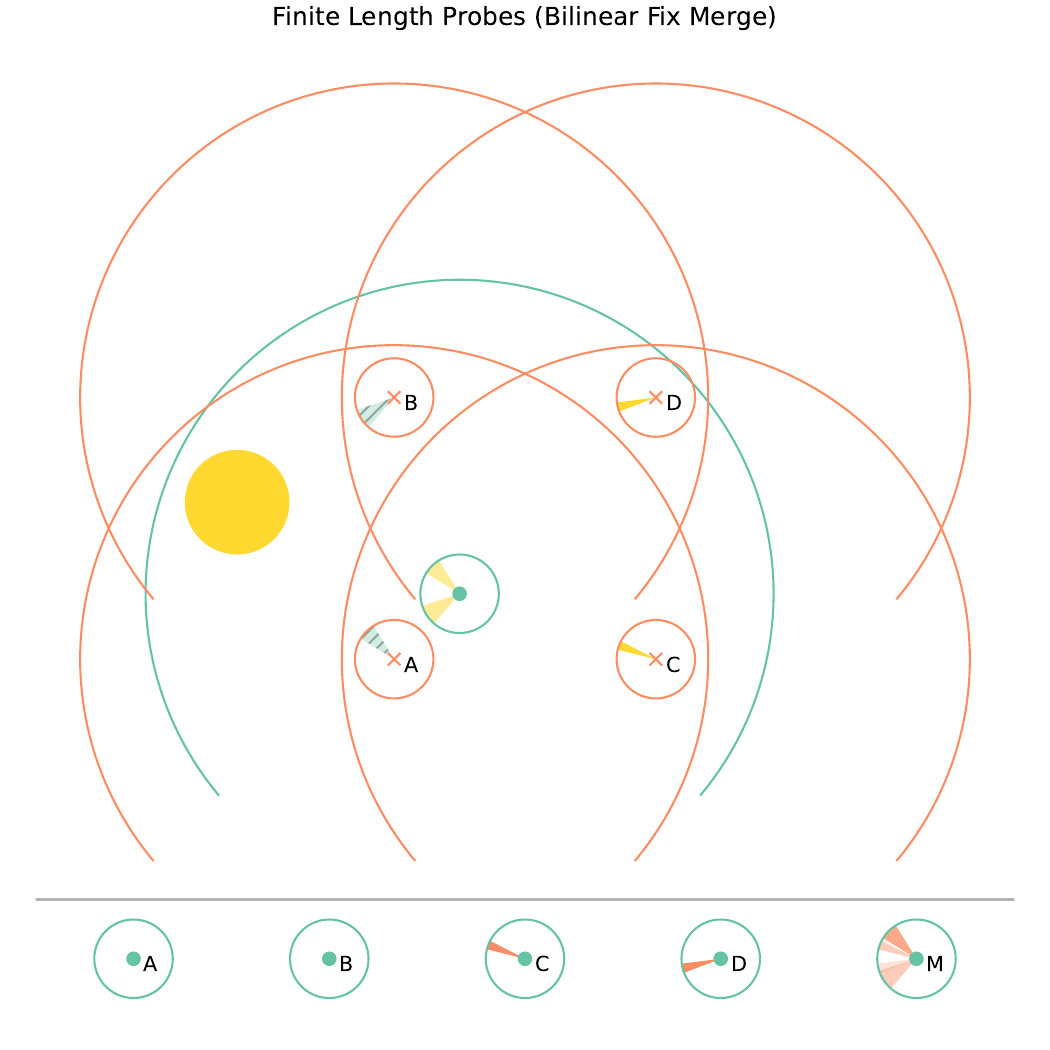}
    \caption{Illustration of the same scenario as Figure~\ref{Fig:ParallaxStandard}, but with the bilinear fix applied. The two contributions on the green probe of cascade $i$ are the result of performing the bilinear fix (tracing rays from their usual start position to the start position of the child rays in cascade $i+1$ for each of the four probes) for probes A and B. The bilinear fix rays from the probe of cascade $i$ to the start of rays from probes C and D do not reach the source. The final merged result M no longer contains an additional contribution (as the bilinear weights of the four contributions correctly sum to unity), and the angular size of the light source is correct (to discretisation error from probe spacing). Whilst there are multiple components from different directions representing the same source, in practice the only occurs over a small range of distances, and the angular spread is dramatically decreased at larger source distances.}
    \label{Fig:ParallaxBilinear}
\end{figure}

The bilinear fix therefore remedies this by effectively reprojecting the sampling rays of each probe to those of the four bilinearly sampled probes of cascade $i+1$, minimising parallax effects, and ensuring occlusion from cascade $i$ is correctly applied.
An illustration of the effects of this fix is shown in Figure~\ref{Fig:ParallaxBilinear}.
This is an extreme case where the radiance interval length is short compared to the probe spacing and the light source is close to the probes, thus exaggerating the parallax effects.
The final merged result, labelled M, contains four different contributions, both from the parallax corrected rays of the cascade $i$ probe to the ray start positions of A and B, but also from C and D. These represent the correct integrated radiance from the light source (to discretisation error, given the probe spacing), thanks to the corrected weighting of each sample through one set of bilinear weights, but the samples are discontinuous in angle due to the extreme nature of this scenario.
In applications with typical parameters, this angular spreading of radiance will be much lower.

\subsection{Discussion}\label{Sec:RcDiscussion}

Building from the penumbra criterion (Section~\ref{Sec:PenumbraCriterion}) we have constructed a framework to exploit the properties of radiation fields to efficiently compute and store an arbitrarily complex field.
We stress that the construction of radiance cascades presented here is neither unique nor necessarily optimal, rather, it is one that is simple to implement.
For example, the grid of probe placements shown in Figure~\ref{Fig:RcGrid} is a simple construction that meets the chosen scaling law, but relies on $n$-linear interpolation.
In a three-dimensional problem, 8 samples of cascade $i+1$ are therefore required for each sample of cascade $i$.
This could potentially be improved upon with a different grid layout allowing for efficient use of a simplex interpolation strategy.

We also note that the use of a radiance cascades framework does not enforce any particular method for the calculation of the radiance intervals.
For example, radiance interval merging can be exploited to compute long-range radiance intervals by a process known as radiance interval extensions.
In this situation, a conventional formal solver is employed to trace an initial small fraction of each interval (at least the distance between the probes) before recursively extending these intervals by merging with the interpolated value at their endpoint.

\section{Implementation Practicalities}

In the following we describe the initial version of \Dex{}, a two-dimensional non-LTE radiative transfer code utilising radiance cascades intended primarily for the modelling of solar isolated structures, such as prominences.
These structures are cool and dense plasma suspended in the hot and tenuous solar corona.
They are the epitome of non-LTE plasma physics typically only visible in narrow spectral lines due to the Sun's radiation field radiatively exciting their component plasma \citep{labrosse_physics_2010}.
Due to the ease of observing these structures in detail from Earth, along with their intensely non-equilibrium nature, cool isolated solar structures are a key tool for developing our understanding of non-equilibrium plasma throughout the universe.
Radiative transfer models of these structures have typically considered simplified atmospheric models, consisting of of one- or two-dimensional slabs of either isothermal and isobaric plasma \citep[e.g.][for 1D and 2D, respectively]{gouttebroze_hydrogen_1993,paletou_two-dimensional_1993}, or with the addition of a prominence-to-corona-transition region (PCTR) obeying functional magnetohydrostatic equations \citep[e.g.][for 1D and 2D respectively]{labrosse_non-lte_2004,heinzel_prominence_2001}.
The history of these models is reviewed in \citet{labrosse_physics_2010}.

Meanwhile, MHD models of prominences, with drastically simplified radiative treatments, have evolved dramatically, investigating mechanisms for the formation and stability of these structures, and including detailed behaviour at high-resolution in two- and three-dimensions \citep[e.g.][]{zhou_simulations_2020,martinez-gomez_transverse_2022,jenkins2022,jercic_prominence_2024}.
There have been few attempts to perform non-LTE synthesis of these models, outside of the use of a simplified \Ha{} proxy \citep{heinzel_fast_2015,gunar_3d_2015}, until an initial attempt to unify the radiative treatments of isolated solar structures with those employed in the lower solar atmosphere by \citet{jenkins_15d_2023}.

The complex structuring of modern prominence models, combined with their high optical depths in some spectral lines, makes them prime candidates for internal shadowing and radiative pumping effects that are poorly considered in the current generation models.
Their fine structure makes the use of short characteristics prohibitively expensive in multiple dimensions \citep{jenkins_bright_2024}.
Additionally, these isolated structures cannot be treated entirely in a vacuum, as recently \citet{jenkins_bright_2024} showed that they may have non-negligible albedo, increasing the radiation field between them and the solar surface, and in turn influencing the lower solar atmosphere, causing significant ionisation.
To this end, we have developed \Dex{} to provide non-LTE radiative transfer in models of solar prominences with much greater accuracy than has previously been possible.

\subsection{2D Invariance}

The radiance cascades examples presented in Section~\ref{Sec:Method} demonstrate a monochromatic formal solution in a flat two-dimensional plane, and this method generalises directly to three dimensions.
To correctly integrate $J$ and the necessary radiative rates it is necessary to integrate over the angular unit sphere at each discretised cell in the volume, so slight modifications to the flatland scheme are needed.
As is convention in two-dimensional radiative transfer, we take the model to be infinite and homogeneous along the axis (henceforth $y$-axis) perpendicular to the two varying axes.
To achieve this we augment the flatland quadrature discussed in Section~\ref{Sec:Method} with a number of inclined rays.
In all examples presented here we use a Gauss-Radau quadrature with eight rays for inclination, so each ray of the flatland case is effectively replaced with 8 rays with the same $x-z$ projection and different inclinations in $y$.
This setup integrates one hemisphere of the unit sphere, which in the case of a homogeneous $y$-axis is symmetric to the other hemisphere.

\subsection{Formal Solution \& Grid Structure}

The \Dex{} formal solution uses a full radiance cascades based solution as described in Section~\ref{Sec:Method} at each wavelength.
The monochromatic radiative transfer equation (RTE) is given by
\begin{equation}
    \hat{\omega}\cdot \nabla I_{\nu,\hat{\omega}} = \eta_{\nu,\hat{\omega}} - \chi_{\nu,\hat{\omega}} I_{\nu,\hat{\omega}},
\end{equation}
where $\eta_{\nu,\hat{\omega}}$ and $\chi_{\nu,\hat{\omega}}$ are the monochromatic emissivity and opacity at frequency $\nu$ in the model from a particular point in a direction $\hat{\omega}$.
To compute $\mathcal{R}_{a,b}$ in \eqref{Eq:Rab} the RTE is solved along a ray by a piecewise constant direct integration.
In this scheme, the ray segment passing through a cell of the domain integrates as if this cell has uniform thermodynamic properties and atomic level populations, i.e. for a path length $\mathop{ds}$ within a cell $i$

\begin{equation}
\begin{cases}
    \tau_\nu(s+\mathop{ds}) &= \tau_\nu(s) + \chi_{\nu, i, \hat{\omega}} \mathop{ds}, \\
    I_\nu(s+\mathop{ds}) &= I_\nu(s) \exp(-\chi_{\nu, i, \hat{\omega}} \mathop{ds}) \\
    &\quad+ \frac{\eta_{\nu, i, \hat{\omega}}}{\chi_{\nu, i, \hat{\omega}}}(1 - \exp(-\chi_{\nu, i, \hat{\omega}} \mathop{ds})),
\end{cases}
\end{equation}

where $\eta_{\nu, i, \hat{\omega}}$ and $\chi_{\nu, i, \hat{\omega}}$ are the emissivity and opacity from all sources at a particular frequency in cell $i$ for the direction aligned with this ray.
This effective definition of the source function $S_\nu = \eta_\nu / \chi_\nu$ differs from the one used in \Lw{} \citep{Osborne2021} and RH \citep{Uitenbroek2001} by the omission of the background scattering term.
In our current models of isolated structures the basic background scattering effects (e.g. Thomson scattering) are ignored.

The formal solution in \Dex{} distinguishes between velocity-dependent and -independent contributions to emissivity and opacity.
At the start of a formal solution for a wavelength, the velocity-independent terms (continua and lines in cells with no bulk velocity) are computed once and stored, whereas velocity-dependent terms (spectral lines) are computed on the fly for each ray given its specific direction, and thus projected plasma velocity passing through a cell.
These calculations are performed in the observer's frame.
Each cell is tagged based on the magnitude of its velocity as to whether velocity-dependent terms are necessary, by default this threshold is 1/2 of the thermal velocity of the atom associated with the governing transition at this wavelength.
Governing transitions are explained in Sec.~\ref{Sec:NonLteImpl}.

In the current work we adopt a ``back-to-front'' solution to the radiative transfer equation.
That is, radiance intervals are computed \emph{towards} their associated probe.
This method works equally well with a ``front-to-back'' solution instead, which could potentially allow for a more efficient solution, allowing early termination from the calculation of long radiance intervals at high optical depths.

As discussed in Section~\ref{Sec:RcDiscussion} the choice of a piecewise constant scheme is effectively arbitrary, and can be changed without affecting the radiance cascade structure.
It is adopted here to simplify interoperation with most MHD codes, which typically follow finite volume definitions.
In the finite volume framework, the values stored per cell represent the average over the volume, rather than a pointwise value, as would be the case in a finite difference method.
Short characteristics methods have typically employed finite-difference solutions, assuming the thermodynamic parameters and atomic energy distribution is known at discrete points in the simulation domain.
The RTE is then solved along characteristics where the evolution of the source function and opacity are assumed to follow a prescribed variation e.g. piecewise parabolic \citet{auer_two-dimensional_1994}, monotonised quadratic Bézier \citep{Stepan2013}, or third order Bézier \citep{DelaCruzRodriguez2013}.
This style of variation is not directly compatible with the approach taken to compute the thermodynamic parameters in MHD codes, and typically requires interpolation to the start and points of each characteristic, introducing an additional potential source of error.

Radiance intervals are spawned relative to the radiance probes of the cascades.
The probes of cascade 0 are placed at the centres of the grid cells.

In \Dex{} only uniform grids are supported due to the current radiance cascades implementation which depends on uniform spacing between probes in a cascade, and also due to the efficiency of traversing such a grid using a digital differential analyser (DDA) strategy \citep[conceptually the same as Bresenham's line drawing algorithm,][]{bresenham_algorithm_1965}.
Additionally, most MHD models used in solar simulations have adopted a uniform grid spacing (potentially with a stretch along one axis), so supporting a general arbitrary grid spacing is unnecessary \citep[e.g.][]{vogler_simulations_2005,gudiksen_stellar_2011,xia_mpi-amrvac_2018,keppens_mpi-amrvac_2023}.
This uniform spacing could be an issue for the formation of very thick lines such as Ly $\alpha$, but if the PCTR is sufficiently resolved in the model, so as to have an exponential variation in density, then as discussed by \citet{mihalas_two-dimensional_1978}, this is a reasonable approach.

\subsection{Non-LTE Iteration}\label{Sec:NonLteImpl}

Aside from the radiance cascade formal solver, \Dex{} adopts a conventional non-LTE setup similar to that of \Lw{} \citep{Osborne2021}.
The preconditioned rate matrix is produced by the ``same-preconditioning'' multi-level accelerated lambda iteration (MALI) method of \citet{Rybicki1992}, instead of the ``full-preconditioning'' approach adopted in \Lw{}.

We have found the ``same-preconditioning'' to often converge slightly faster than the ``full-preconditioning'' approach.
Additionally, it requires less memory due to the elimination of the cross-coupling terms.
Whilst these aren't required to be stored in the ``full-preconditioning'' approach, there is a substantial computational cost to not doing so, requiring additional loops across all overlapping transitions.

For simplicity, a diagonal approximate lambda operator (ALO) $\Lambda^*$ is adopted, and due to the piecewise constant formulation of the formal solver the monochromatic, directional ALO is given by
\begin{equation}
    \Lambda^*_{\nu,\hat{\omega}}(k) = \exp{\left( -\tau_{\nu,k,\hat{\omega}} \right)},
\end{equation}
where $\tau_{\nu,k,\hat{\omega}}=\chi_{\nu,i,\hat{\omega}} \mathop{ds}$ is the monochromatic optical depth from the centre to the edge of cell $k$ along direction $\hat{\omega}$.
As usual, the $\Psi^*_{\nu,\hat{\omega}}$ operator required by the MALI method is computed from $\Lambda^*_{\nu,\hat{\omega}}(k) / \chi_{\nu,k,\hat{\omega}}$.
Due to the purely local nature of the ALO, it is only computed for cascade 0, and temporarily stored before a second pass that accumulates the terms in the rate matrix $\Gamma$ from this wavelength.

The numerical implementation of this MALI method is similar to that of \Lw{} detailed in appendix B of \citet{Osborne2021}.
Using the notation therein the $\mathcal{U}_j$ term of (B17) is then replaced by $U_{ji}$ for the transition in question (with upper and lower level $j$ and $i$ respectively) and $\mathcal{X}_l$ of (B18) is replaced by $\chi^\dagger_{ij}$, removing the two summations necessary for the cross-coupling terms.
The effective intensity term $I^\mathrm{eff}$ (B19) is now per transition rather than per atom and computed as
\begin{equation}
    I^\mathrm{eff}_{\nu,\vec{d};\,ji} = I^\dagger(\nu,\vec{d}) - \Psi^*_{\nu,\vec{d}}U_{ji},
\end{equation}
in the notation therein, or
\begin{equation}
    I^\mathrm{eff}_{\nu,\hat{\omega};\,ji}(k) = I^\dagger_{\nu, \hat{\omega}}(k) - \Psi^*_{\nu, \hat{\omega}}(k) \cdot U_{ji}(k)
\end{equation}
in the notation of this work, where $\dagger$ refers to the value at the previous value of the level populations, in the case of intensity representing the value just calculated inside the formal solver.
As in \Lw{}, only the off-diagonal terms of the rate matrix $\Gamma$ are accumulated during each iteration, whilst the diagonal terms is filled in by the conservation equation requiring that the total transition rate out of each level match the sum of transitions from this level into every other level.
The $\Gamma$ matrix is then used to update the populations towards statistical equilibrium by solving
\begin{equation}
    \Gamma_s \vec{n}_s = \vec{0}
\end{equation}
for each species $s$.
No forms of convergence acceleration are currently employed.

Charge conservation can also be enforced via a single Newton-Raphson iteration after each statistical equilibrium update as originally described by \citet{Heinzel1995} and \citet{paletou_two-dimensional_1995}.
The method employed here is the same as the generalisation to MALI with overlapping transitions presented in Appendix D of \citet{Osborne2021}, but currently only includes hydrogen in the Newton-Raphson matrix.
This approach is also employed to determine a self-consistent electron density at the start of a simulation by considering only the collisional rates (thus setting the populations to LTE).

As numerical models of prominences are often specified in terms of pressure, we also allow for pressure conservation (requiring charge conservation to be enabled).
This approach scales the total hydrogen number density (from which all other atomic populations are scaled) such that
\begin{equation}
    P = N_{\mathrm{tot}} k_B T = (A_{\mathrm{tot}} n_{H\mathrm{tot}} + n_e) k_B T
\end{equation}
is held constant to a value provided with the input atmosphere.
Here $N_\mathrm{tot}$ is the total number density of atoms and electrons, $A_\mathrm{tot}$ is the total abundance of atomic matter relative to hydrogen, $n_{H\mathrm{tot}}$ is the hydrogen number density, $n_e$ is the electron number density, $k_B$ is the Boltzmann's constant, and $T$ the local plasma temperature (assuming a one-fluid model).
The update to $n_{H\mathrm{tot}}$ is performed by simply computing the difference in pressure due to the change in electron density from charge conservation, and adjusting $n_{H\mathrm{tot}}$ to compensate.
Whilst this approach assumes a linear problem, this additional iteration converges quickly to a very low change per iteration.

Spectral lines are assumed to be Voigtian, and for the Voigt profiles, we compute a large lookup table (by default 256\,MB), which is bilinearly interpolated.
This table is filled using the approximation of \citet{humlicek_optimized_1982}.

\Dex{} is designed to support multiple concurrent atomic models with overlapping transitions.
We consider this to be an essential requirement for modern radiative transfer frameworks, due to significant interactions between radiative transitions that otherwise need to be treated manually, and could potentially be overlooked \citep[e.g. for the photoionisation of the \Caii{} by the Lyman continua and lines,][]{ishizawa_emission-line_1971,gouttebroze_calcium_2002,osborne_importance_2021}.

Whilst the wavelength integration of the rate matrix in \Dex{} is the same as \Lw{}, the wavelength grid is created differently.
The wavelength range is divided into bands wherein a governing transition is selected.
A transition becomes governing by
\begin{itemize}
    \item Being the spectral line closest to its rest wavelength of all active spectral lines,
    \item Being the only spectral line active at a wavelength,
    \item Being the active continuum closest to its edge,
\end{itemize}
where \emph{active} refers to being inside the range of wavelengths where a transition is said to participate.
The wavelength grid is then made by merging the wavelength grids of each transition over the ranges where it is the governing transition.
This prevents the overly fine and irregular wavelength spacing that can often occur when merging overlapping wavelength grids in \Lw{}.
This method of merging is similar to the final step of frequency mesh generation of \citet{castor_new_1992}.

\subsection{Boundary Conditions}

\Dex{} supports two boundary conditions:
\begin{enumerate}
    \item A simple free-space with no incoming radiation.
    \item A tabulated form of the \Pw{} boundary condition as described in \citet{jenkins_15d_2023} and \citet{peat_doppler_2024}.
\end{enumerate}
The latter of these is a table of emergent intensity from an arbitrary plane-parallel atmosphere as a function of inclination and wavelength, computed by \Lw{}.
By default the FAL C model of \citet{fontenla_energy_1993} is adopted.
On the highest cascade, rays that start outside the volume are checked for intersection with a sphere of solar radius, and sample intensity based on their angle to the normal at intersection.
Due to the very high angular resolution on the highest cascade, we have not found it necessary to perform the additional boundary condition samples detailed in \citet{jenkins_15d_2023} to correct the integral over the angular quadrature.
This model currently does not include coronal extreme ultraviolet emission for photoionisation, but could easily be extended to support it.

\subsection{Sparsity}\label{Sec:Sparsity}

In swept methods, such as the short characteristics approach typically adopted in solar radiative transfer, the radiation field is carried downwind across the domain by the formal solver.
Thus, even if a cell can be proven to not interact with the radiation at a given frequency (for example a chromospheric spectral line and coronal temperatures and densities), it must still be involved in the sweep.
As our radiance cascades implementation is more akin to a long characteristics method, we can first determine the cells of interest for a particular non-LTE problem.
In the current implementation this is simply a temperature flag, typically set to \SI{250}{\kilo\kelvin}.
This flag then defines the probes of cascade 0 that may be involved in the non-LTE problem.
Their bilinear ``parents'' are recursively flagged up to the maximum cascade, and the radiance intervals are only computed for these flagged probes, and non-flagged cells are treated as transparent with no emission.
This can lead to dramatic reductions in computational cost for models with complex structure but where a large fraction of the domain is non-participating media (such as corona).
In this implementation the memory utilisation is the same whether the sparse solution is employed or not, however, the simple grid currently employed could be replaced with an octree or VDB-like structure \citep{museth_vdb_2013,museth_nanovdb_2021}, allowing for reductions in memory consumption, whilst also accelerating the traversal of long-range radiance intervals through free-space.

\subsection{Parallelisation, single precision \& GPUs}

When computing a radiance cascade each radiance interval can be computed entirely independently -- allowing all radiance intervals of a single wavelength to be evaluated in parallel.
Unlike the commonly employed short characteristics solvers, radiance intervals can be long, and span a significant portion of the domain.
Multi-process parallelisation has not been implemented in the current version of \Dex{}, but is in development to support larger models.

\Dex{} is written in performance-portable C++20 using the YAKL framework \citep{norman_portable_2023}, and makes use of C++ metaprogramming features such as templates to allow for generation of efficient functions based on different input parameters.
The use of YAKL allows \Dex{} to be compiled for different system architectures, both parallel Central Processing Units (CPUs) and NVIDIA, AMD, and Intel Graphical Processing Units (GPUs).
Currently \Dex{} is accelerated by the use of single GPUs where the entire model can fit in memory.
This style of embarrassingly-parallel computation maps well to GPUs, so long as care is taken to minimise divergence between the threads of a block.
GPUs operate threads in groups\footnote{These are referred to as warps on NVIDIA hardware, warps or wavefronts on AMD, and subgroups on Intel. These typically take values $\sim$32.} similar to the vector units of CPUs and operate significantly faster when each group makes contiguous memory access and takes the same branch at conditionals.
Within the radiance cascade computation, the majority of processing time is spent solving the radiative transfer equation along each radiance interval (primarily computing the directional contributions the emissivity and opacity for moving models).
To optimise for this, the cascades in \Dex{} are laid out so that each block of inclination rays is contiguous, and the remaining threads of the block are filled with the equivalent paths for different wavelengths.
This minimises divergence among threads of a block, but can be disabled to reduce memory cost in large models.

Consumer GPUs, such as those typically employed for machine learning, perform single precision floating point operations \emph{much} faster than double precision, with the typical ratio for current NVIDIA offerings being 1:64.
With careful selection of units, the entire formal solution and construction of the rate matrix can be performed in single precision.
To this end, \Dex{} considers specific intensity in units of \si{\kilo\watt\per\square\metre\per\nano\metre\per\steradian}.
The direct solution of the rate matrix as per \citet{Rybicki1992} is not reliably stable in single precision, however if the diagonal of the matrix (to ensure population conservation) is recomputed in double precision, then the matrix solution in double precision yields the expected result.

The radiance cascades method highlights the effectiveness of GPU acceleration for this style of calculation.
On a workstation with an Intel i7-13700 CPU and an NVIDIA RTX 4070 GPU, the GPU performs the formal solution $80\times$ faster, while consuming $2\times$ the power.
This highlights the importance of employing GPU-based solutions ($40\times$ lower energy consumption here) to reduce the environmental impact and improve the equitability of radiative transfer calculations.
To this end, \Dex{} does not currently support solving the population update from the rate matrix on CPU (however single formal solutions are supported).
On GPU the parallel batched solution to the system of linear equations resulting from the rate matrix at each point in the domain is provided by the MAGMA library\footnote{\url{https://icl.utk.edu/magma/}} \citep{dongarra_accelerating_2014}.

\subsection{Input/Output}

\Dex{} is a standalone C++ program, and operates on two different kinds of files:
\begin{itemize}
    \item Simple human-readable configuration files in the YAML format for simulation configuration and atomic model data.
    \item Binary storage of array data for input atmosphere, boundary conditions, and output as netCDF 4\footnote{netCDF is developed by UCAR/Unidata (\url{http://doi.org/10.5065/D6H70CW6}).}.
\end{itemize}

A pydantic\footnote{\url{https://pydantic.dev/}} schema is available for each of the YAML files, allowing for easy manipulation and validation of configuration values in Python before running a model.
The YAML-based format for atomic data (Common Radiative Transfer Atomic Format -- CRTAF) is a draft of a tentative standard for unifying atomic models for use in radiative transfer codes, and has an associated specification\footnote{\url{https://github.com/Goobley/CommonRTAtomicFormat}} and Python package\footnote{\url{https://github.com/Goobley/crtaf-py}}, currently capable of interoperating with \Lw{} model atoms.

\subsection{Modes}

Several modes of operation are available in \Dex{}
\begin{itemize}
    \item \texttt{GivenFs}: the input model provides a set of isotropic emissivities and opacities at each point in the domain. These are solved for $J$, allowing for the calculation of \emph{ad hoc} models such as Figure.~\ref{Fig:RcVsSc}.
    \item \texttt{Lte}: An input atmosphere and one or more atomic models are provided. \Dex{} computes $J$ throughout the model assuming LTE atomic level populations and potential boundary conditions.
    \item \texttt{NonLte}: The full non-LTE iterated solution is computed for the provided atmosphere and atomic models, optionally performing pressure and charge conservation as detailed in Section~\ref{Sec:NonLteImpl}.
\end{itemize}

\subsection{Spectral Output}\label{Sec:OutputSpectra}

In conventional short characteristics codes, the emergent intensity along the rays of the quadrature can be output from the formal solution during the iterative process.
This quantity is not directly available from the radiance cascades method employed in \Dex{}.
To this end, we also provide a simple tool for post-processing the output of \Dex{} models: a single-pass long characteristics formal solver, \texttt{dexrt\_ray}, for rays of arbitrary direction through the domain.
This tool can also output line formation diagnostics including the contribution function at each point along each ray's path.
As discussed by \citet{de_vicente_long_2021}, a long characteristics formal solution provides a notably spatially sharper solution than short characteristics for inclined rays.

The single-pass nature of this tool utilises a ``front-to-back'' formulation of the radiative transfer equation: each ray is traced from the point where the emergent intensity is to be found, accumulating opacity and attenuated emission from each cell traversed.
All key spectral diagnostics, including the contribution function, can be cast in a form computable with this approach.
The efficiency of this single-pass solution allows for calculation and output of a handful of wavelength points from one viewpoint at close to interactive framerates.
This tool could additionally be repurposed for considering off-axis rays through grids of atomic populations computed using 1.5D plane-parallel methods.

\section{Application}

To demonstrate the effectiveness of \Dex{} for models of solar prominences we apply it to a snapshot of a 2.5D MHD model of a solar prominence computed by \citet{Jenkins2021}.
This model is computed with MPI-AMRVAC \citep{xia_mpi-amrvac_2018,keppens_mpi-amrvac_2023}, leveraging the adaptive mesh refinement capabilities of this code.
It studies the self-consistent in-situ formation of a solar prominence by radiative cooling and condensation of coronal plasma.
The volume is initialised with a linear force-free field, and a flux rope is created via converging and shearing footpoint motions applied to the lower boundary.
Pockets of increased density form, and are subject to both a runaway thermal instability (due to the optically thin radiative losses), and Rayleigh-Taylor instability as they gravitationally slip through surrounding less dense material.
These fall until supported by the magnetic field to form a primary condensation which continues to cool and grow via thermal instability.

In Section~\ref{Sec:Snapshot} and \ref{Sec:NonLteSetup} we describe the setup of our model, before presenting spectral results in Section~\ref{Sec:Spectra} and comparing with the commonly employed \Ha{} proxy of \citet{heinzel_fast_2015} in Section~\ref{Sec:HaProxy}.
Then, in Section~\ref{Sec:LybFormation} we investigate the formation of \Lyb{} line in this model.
Finally, we discuss the performance and parameters of the \Dex{} solution in Section~\ref{Sec:DexPerformance}.
The analysis of this snapshot is not intended to be exhaustive, but instead motivate the use of modern multidimensional radiative transfer tools such as \Dex{} for this style of model.

\subsection{MHD Snapshot}\label{Sec:Snapshot}

\begin{figure*}
	\centering
    \includegraphics[width=\textwidth]{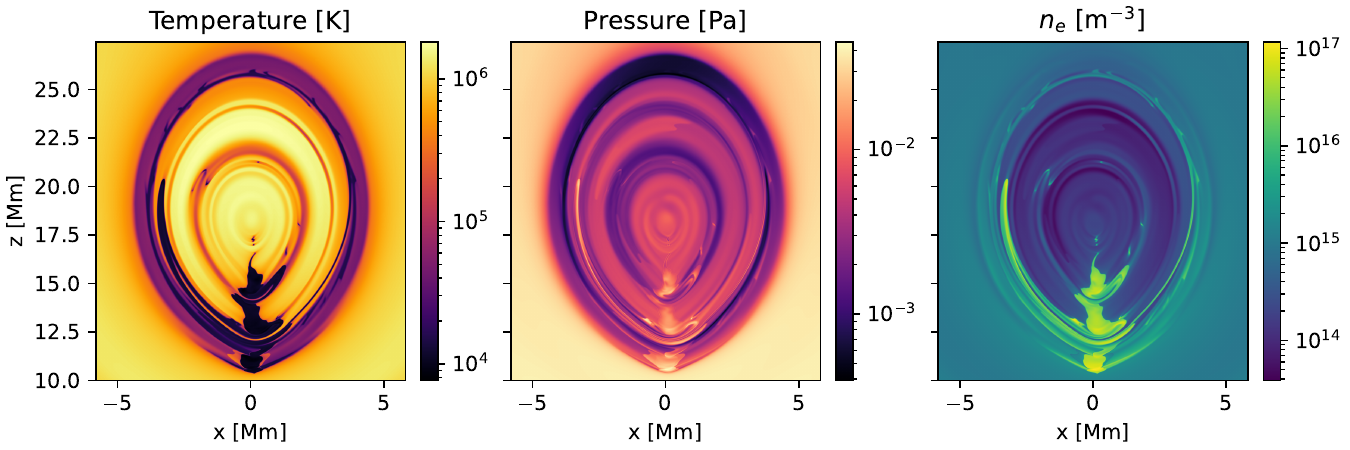}
    \caption{Parameters of the MHD snapshot ($t=\SI{4519}{\second}$ of the high-resolution \SI{3}{\gauss} model of \citet{Jenkins2021}) used for our non-LTE model. The temperature and pressure are extracted directly from the MPI-AMRVAC model, whilst the electron density ($n_e$) is a result of the charge conservation applied during iteration with \Dex{}. The $z$-axis is taken as the normal to the solar surface at $x=\SI{0}{\mega\metre}$, whilst the $x$-axis is perpendicular to this, and tangent to the solar surface.}
    \label{Fig:ModelParams}
\end{figure*}

One of the high-resolution snapshots (with an effective resolution of \SI{5.7}{\kilo\metre}) analysed in \citet{Jenkins2021} was selected as input to our non-LTE model.
Due to presenting the most complex temperature and density structure the $t=\SI{4519}{\second}$ snapshot of the \SI{3}{\gauss} model was chosen.
The temperature and pressure of this snapshot are shown in Figure~\ref{Fig:ModelParams}, alongside the electron density obtained through charge and pressure conservation.
The $z$-axis is taken as normal to the solar surface at $x=\SI{0}{\mega\metre}$, whilst the $x$-axis is perpendicular to this, and tangent to the solar surface.
As such, a view down the $z$-axis will see the model backlit by the solar disk as a filament, and along the $x$-axis as a prominence.
The flux rope contains a highly stratified temperature and density structure, with a multi-layered primary condensation in the lower centre of the domain.
There is a notable ``arm'' condensation feature in the $-x$ region stretching just past an altitude of $z=\SI{20}{\mega\metre}$.

This model, slightly cropped from the full domain of the original simulation, has an $x \times z$ resolution of $2048 \times 3072$ with a voxel scale of \SI{5.7}{\kilo\metre}.

\subsection{Non-LTE Model}\label{Sec:NonLteSetup}

We apply \Dex{} to this snapshot to solve for the self-consistent radiation field, statistical equilibrium populations, and ionisation state of hydrogen and calcium.
These two elements are commonly used in prominence observations and their transitions are commonly used to probe different layers of atmospheric models \citep{labrosse_physics_2010}.
Classic atomic models are adopted for these two elements:
\begin{itemize}
    \item A five-level plus continuum model for hydrogen with ten lines following the approach of \citet{johnson_approximations_1972}. This model is the default model for hydrogen employed in \Lw{} and RH.
    \item A five-level plus continuum model for \Caii{} with five lines (H, K, and the infrared triplet), where the continuum level represents Ca\,\textsc{iii}. This model is the default model for \Caii{} employed in \Lw{} and RH.
\end{itemize}
Both charge and pressure conservation, as described in Section~\ref{Sec:NonLteImpl} are employed.
Partial frequency redistribution (PRD) effects are not considered.

The \Pw{} boundary condition, tabulated by \Lw{}, is used, ensuring self-consistently diluted and limb-darkened (as computed from a plane-parallel model) radiation is included at every wavelength, containing irradiation from these same two atomic models, in addition to LTE background terms from other species.
After merging the wavelength grids, 709 wavelength points remain in the model; these models are essentially unchanged from those used in typical plane-parallel models, without the common reduction in wavelength quadrature points.

Using a cutoff temperature of \SI{250}{\kilo\kelvin}, as discussed in Section~\ref{Sec:Sparsity}, approximately 25\% of the cells are flagged as active, with an equivalent number of the probes of cascade 0, and an increasing fraction of the upper cascades to approximately 33\% of the uppermost cascade.

Our convergence criterion for the iteration is the maximum relative difference of level populations, electron density, and pressure conservation modifications between iterations dropping below $10^{-3}$.
In practice, this is dominated by the level populations.

\subsubsection{\Pw{} Model}

For comparison, two sets of \Pw{}\footnote{\Pw{} is a Python package using \Lw{} to generate plane-parallel spectra of isolated solar structures such as prominences.} models were also computed.
These use the same configuration as above, but consider the domain in two different orientations to produce the prominence and filament views, as in \citet{jenkins_15d_2023} and \citet{peat_doppler_2024}.
These models are 1.5D plane-parallel, so neighbouring columns do not interact with each other.
Additionally, there is no guarantee of self-consistency between the same grid cell in a column computed in ``prominence-mode'' to that same cell in a column computed in ``filament-mode''.
In both cases every 16th column was computed, so these models appear somewhat more ``blocky''.

\subsection{Spectra}\label{Sec:Spectra}

\begin{figure*}
	\centering
    \includegraphics[width=\textwidth]{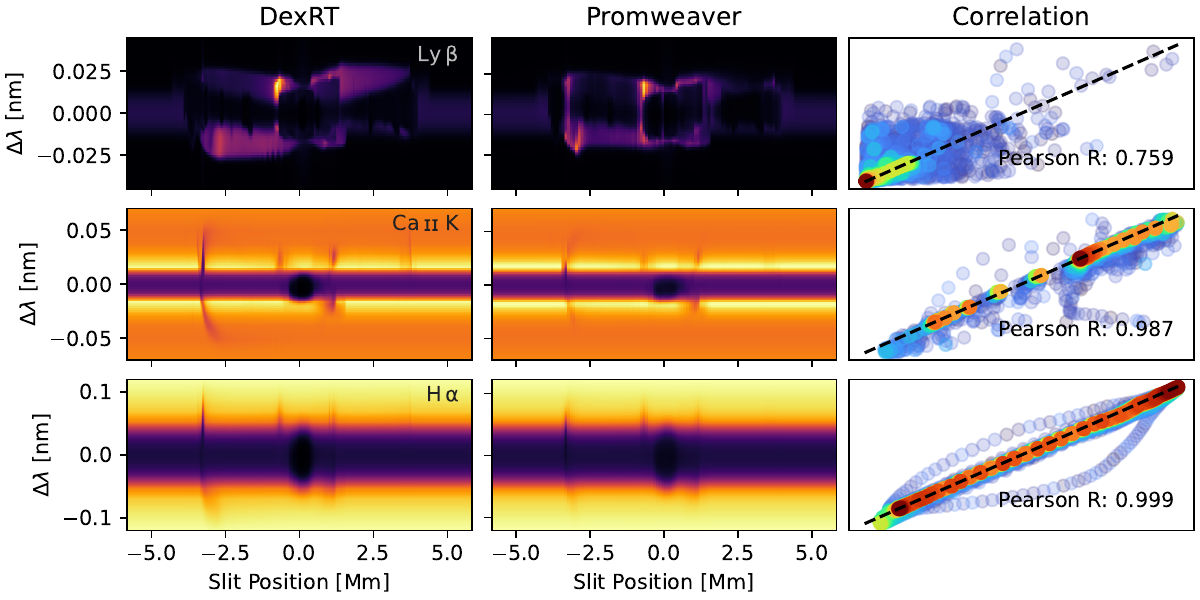}
    \caption{Synthesised spectra for \Lyb{}, \CaK{}, and \Ha{} looking vertically through the model as a filament. The first column shows the spectra obtained with \Dex{}, the second the spectrum from \Pw{}, and the third the correlation between the two. The slit position is analogous to the $x$-axis of the atmospheric model. Each point on the correlation plot is coloured by the density of points in this region.}
    \label{Fig:FilSpectra}
\end{figure*}

\begin{figure*}
	\centering
    \includegraphics[width=\textwidth]{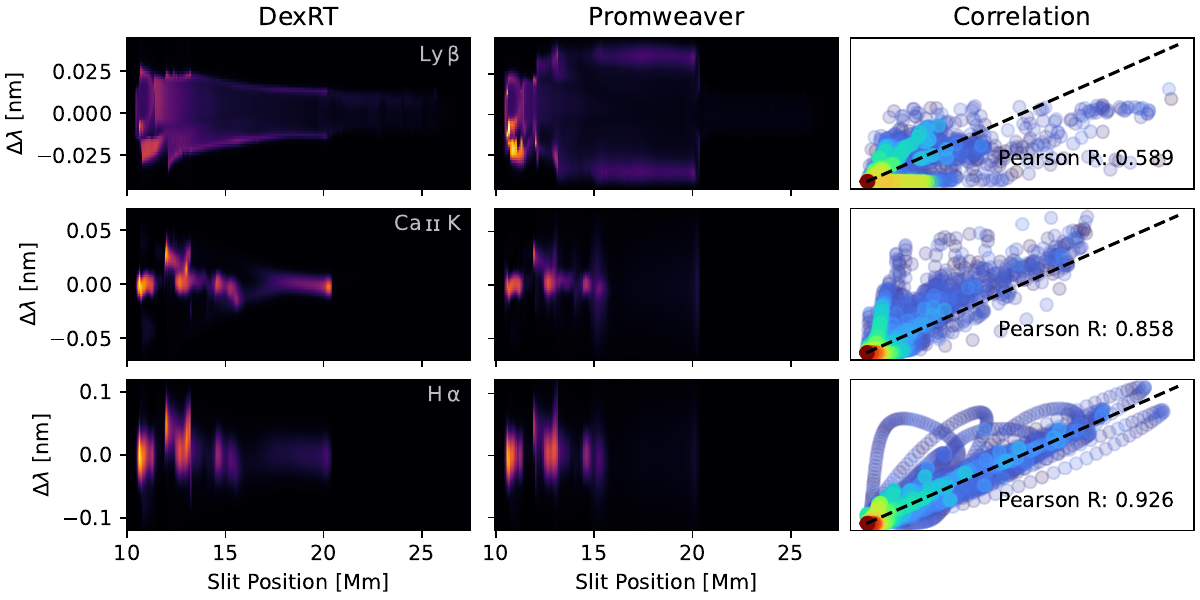}
    \caption{Synthesised spectra for \Lyb{}, \CaK{}, and \Ha{} looking horizontally through the model as a prominence. The first column shows the spectra obtained with \Dex{}, the second the spectrum from \Pw{}, and the third the correlation between the two. The slit position is analogous to the $z$-axis of the atmospheric model. Each point on the correlation plot is coloured by the density of points in this region}
    \label{Fig:PromSpectra}
\end{figure*}

The spectra of the \Lyb{}, \CaK{}, and \Ha{} spectral lines obtained by post-processing with \texttt{dexrt\_ray} (see Section~\ref{Sec:OutputSpectra}) and with \Pw{} for the model in a filament and prominence view are shown in Figures~\ref{Fig:FilSpectra} and \ref{Fig:PromSpectra} respectively.
These figures also include a plot of the correlation between the \Dex{} and \Pw{} spectra.
For each of these spectral lines, a video of the spectrum as a function of viewing angle from the \Dex{} model is available online.

For each of the spectral lines shown in filament view in Fig~\ref{Fig:FilSpectra}, we see a structured spectrum overlying a uniform background spectrum (from the solar disk).
In these panels, the slit position is analogous to the $x$-axis of the input atmosphere.
There is good qualitative agreement between \Dex{} and \Pw{} for the \CaK{} and \Ha{} spectral lines, and this is confirmed by the strong correlation.
We note that the absorption features in the line core from the primary condensation (around slit position \SI{0}{\mega\m}) are slightly darker in the \Dex{} solution.
This is also true for the lighter absorption features on either side of the central condensation in \CaK{}.
The absorption feature corresponding to the ``arm'' at a slit position of \SI{-3}{\mega\m} is far narrower in wavelength and less pronounced in the \Pw{} \CaK{} solution.

The agreement in the \Lyb{} line is much less good.
For slit positions in the $[-4, 4]\,\si{\mega\m}$ range the \Dex{} model predicts three distinct absorption structures along the line core, while the \Pw{} solution presents much more fragmentary absorption effects.
Additionally, the asymmetries (blue for negative, red for positive slit position) and variations in line widths are far more pronounced in the \Dex{} model.
This is particularly evident around $\Delta\lambda=\SI{-0.02}{\nano\m}$ in the $[-3, 0.5]\,\si{\mega\m}$ range, and around $\Delta\lambda=\SI{0.02}{\nano\m}$ in the $[0.5, 3]\,\si{\mega\m}$ range.
The correlation plot shows a large scatter of values at lower intensities with no clear structure.

The effect of the structure creating a strong emission feature in \Lyb{} around slit position \SI{-0.5}{\mega\metre} in Figure~\ref{Fig:FilSpectra} spreads over a much wider range of columns, and is a clear signature of two-dimensional transfer effects modifying nearby populations.

These same spectral lines are shown for the prominence viewpoint in Figure~\ref{Fig:PromSpectra}.
Here, the slit position is analogous to the $z$-axis of the input atmosphere.
The \Dex{} and \Pw{} spectra agree significantly less well than in the filament case, also highlighted by the increased scatter on the correlation plots.
When viewed as a filament there is no background illumination from the solar disk thus all spectral lines are in emission, but may contain absorption features such as central reversals.
In all spectral lines, no significant emission is seen above $\sim\SI{20}{\mega\m}$.
The most striking difference between the \Dex{} and \Pw{} models is the extreme width of the spectral lines for slit positions in the $[16, 20]\,\si{\mega\m}$ range in the \Pw{} model, that isn't present for the \Dex{} treatment.
Emission in this region appears to be mostly due to the ``arm'' structure of the atmosphere.
This effect is most apparent in the \Lyb{} line where a gradual tapering of line width is seen over the $[12,20]\,\si{\mega\m}$ range in the \Dex{} model versus a mostly constant line width in the \Pw{} model over this same range.
The spatial variations in intensity of the \Lyb{} line vary much more smoothly in the \Dex{} model, although there remains complex finely-structured line profiles in the $[10,12]\,\si{\mega\m}$ region where we are observing the bottom of the primary condensation.

Considering now the \CaK{} and \Ha{} lines, we see good agreement in line shapes over the $[10,15]\,\si{\mega\m}$ range corresponding to the bulk of the primary condensation.
Within these structures, the peak intensities are higher in the \Dex{} model, and the asymmetries are more significant.
The spatial intensity falloff from the bright band at the bottom of each separate structure in the $[12,15]\,\si{\mega\m}$ range is also slower in the \Dex{} model, likely a sign of two-dimensional transfer effects.

As previously discussed for \Lyb{}, the differences over the $[15,20]\,\si{\mega\m}$ region corresponding to the ``arm'' structure are dramatic in the \CaK{} and \Ha{} lines.
There is close to no emission in this region in the \Pw{} model, compared to a gradually brightening structure in the \Dex{} solution.
The bright band at \SI{20}{\mega\m} in the \Dex{} output is faintly visible in the \Pw{} spectra.

We note that the reduction in Pearson correlation coefficient in the prominence case will be somewhat due to the lack of consistent (but wavelength-varying) background spectrum present in the filament model.
Nevertheless, the agreement between the full \Dex{} solution, and the 1.5D plane-parallel \Pw{} model is significantly less good for a prominence synthesis than a filament one.
The \Pw{} model is also incapable of synthesising inclined rays through the model, instead simply producing the spectrum along inclined rays through each plane-parallel column.
Videos of these spectra at different viewing angles through the \Dex{} model are available the online supplementary material.

\subsubsection{\Ha{} Proxy}\label{Sec:HaProxy}

\begin{figure*}
	\centering
    \includegraphics[width=\textwidth]{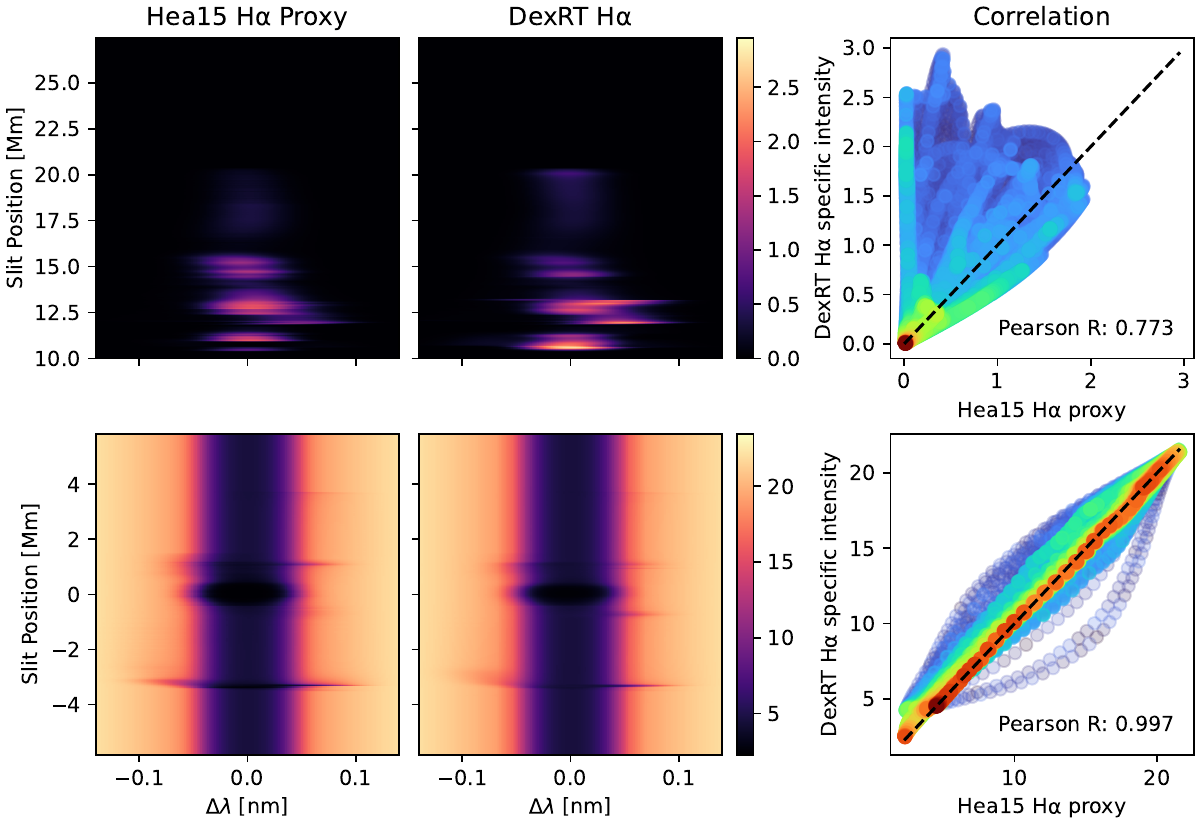}
    \caption{Comparison of the \Ha{} spectrum from \Dex{} from the axis-aligned prominence (upper row) and filament (lower row) viewpoints with the predicted intensity from the \Ha{} proxy of \citet{heinzel_fast_2015}, labelled Hea15. The correlation in the third column shows fair agreement. For the filament projection, the underlying solar spectrum follows the same model is the same in both cases for consistency.
    Specific intensity in units of \si{\kilo\watt\per\square\metre\per\nano\metre\per\steradian}.}
    \label{Fig:Hea15ProxyComparison}
\end{figure*}

The approximate \Ha{} synthesis method of \citet{heinzel_fast_2015} has become a routine tool synthesising emission from complex two- and three-dimensional models \citep[e.g.][]{gunar_3d_2015,claes_thermal_2020,zhou_simulations_2020,martinez-gomez_transverse_2022,jenkins2022}.
The recipe is simple to apply and computationally affordable, although intended for optically thin structures.
It was previously benchmarked against 1.5D synthesis of MHD models by \citet{jenkins_15d_2023}, and found to provide overall good agreement, especially for the columns with total opacity $\tau_\nu<2$ in the \Ha{} line core, with reasonable agreement for more optically thick columns.

In Figure~\ref{Fig:Hea15ProxyComparison} we compare the results of this proxy (labelled Hea15), with \Dex{} for the \Ha{} line, when viewed as a prominence and as a filament.
For consistency, the background for the filament case uses the same spectrum.
The maximum optical depth in the \Ha{} line core is $\sim1.8$ in prominence view and $\sim3.8$ in filament projection.
Our presentation of the proxy differs from most applications due to calculating spectra rather than simply line-core intensities and we assume a Gaussian line absorption profile.

As in the previous section, the agreement between the two treatments when considering a filament is very good.
The \Ha{} proxy creates darker and broader absorption features, visible around $-3$, $1$, and $\SI{4}{\mega\m}$, meanwhile \Dex{} produces a stronger absorption feature in the red wing around \SI{-0.5}{\mega\m}.
The overall agreement between these approaches is very good; note that the scatter plot only includes points with a filament optical depth above 0.01, to limit the effects of the common background.

In the prominence view the limitations of the Gaussian line profile adopted become more evident.
The overall wavelength structure of the two techniques agrees well, however, the \Ha{} proxy is unable to capture the brightenings around $11$, $12$ and $\SI{13}{\mega\m}$, and fails to show the strong wavelength shear associated with the last of these, along with the darkened region above it.
Once again, the ``arm'' structure in the $[16,20]\,\si{\mega\m}$ range differs between treatments, however the \Ha{} proxy, which is based on two-dimensional models, handles this better than the 1.5D \Pw{} model.
The diffuse radiation around \SI{17.5}{\mega\m} is comparable, although the proxy fails to produce the brightening around \SI{20}{\mega\m}.

Whilst not a replacement for the full non-LTE treatment, the \Ha{} proxy of \citet{heinzel_fast_2015} performs admirably in capturing the qualitative structure of this model, but performs notably better on the filament view than the prominence view, despite its higher optical depth.
This is likely related to the fact that the background absorption in each column is more independent for the filament view, whereas the prominence emission will be more greatly affected by $J$ throughout the prominence structure.

\subsection{\Lyb{} Formation}
\label{Sec:LybFormation}

\begin{figure}
	\centering
    \includegraphics[width=0.95\columnwidth]{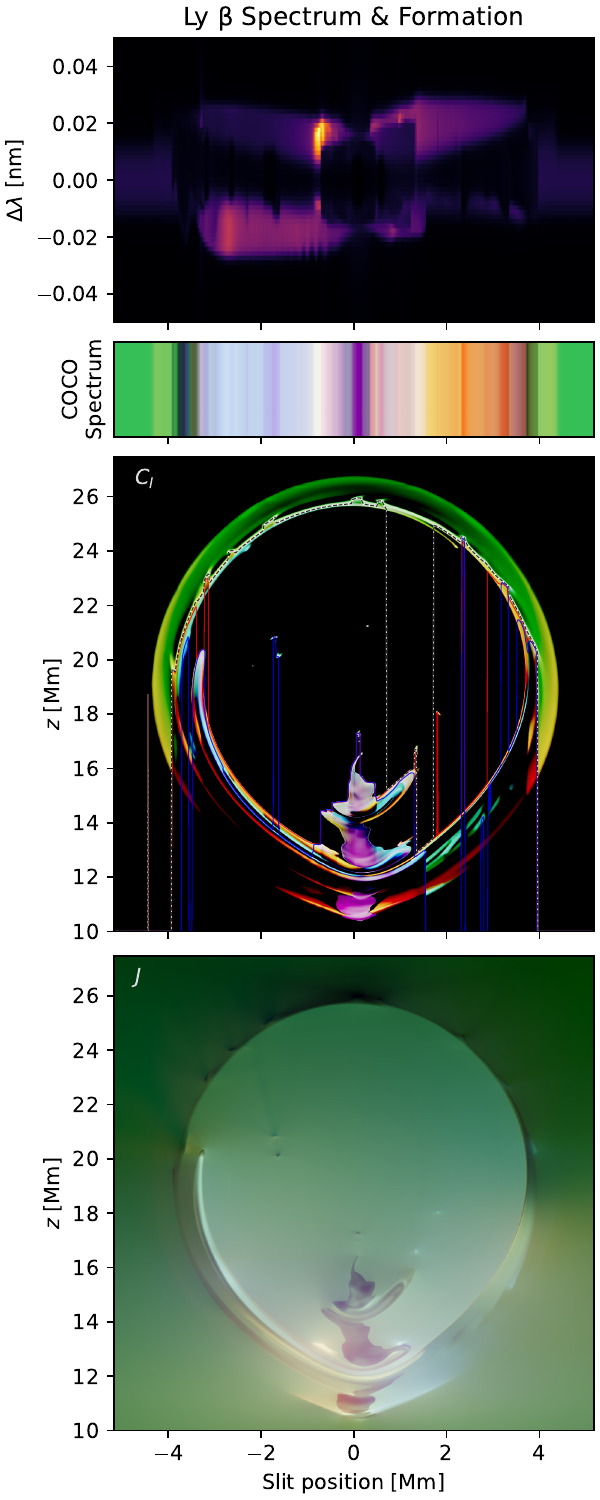}
    \caption{From top to bottom: 1. Ly $\beta$ spectrum when viewed as a prominence. 2. The previous spectrum colour-compressed to highlight line asymmetries. 3. The colour-collapsed contribution function for this line, with the associated weighted $\tau_\nu=1$ for the red wing, blue wing, and line core, in red, blue and white with black dashes respectively. 4. Colour-collapsed $J$ throughout the model. A colourblind-friendly variant of the lower two panels (with the colour channels exploded) is provided as Figure~\ref{Fig:LybColourblind}.}
    \label{Fig:LybFormation}
\end{figure}

The \Lyb{} filament spectrum shown in Figure~\ref{Fig:FilSpectra} appears to be made of two components: a bright underlying structure, with an attenuating absorption layer on top.
This is particularly evident when viewing the video in supplementary material as there is a clear parallax shift between these two layers as the viewing angle rotates.

To investigate the formation of this line \texttt{dexrt\_ray} is used to compute the contribution function (the integrand of the radiative transfer equation, highlighting the contributions to the emergent spectrum).
This is shown, along with the line profile, and $J_\nu$ in Figure~\ref{Fig:LybFormation}.
The colour-collapsed (COCO) methodology adopted in this figure is the same as \citet{druett_cocoplot_2022}, although a tonemapping function is applied to improve the visibility of details in the high-dynamic range data as in \citet{Osborne2022}.
From top to bottom these panels show:
\begin{enumerate}
    \item The \Lyb{} spectrum.
    \item The colour-collapsed spectrum, highlighting how asymmetries correspond to a colour.
    \item The colour-collapsed contribution function, overlaid with the $\tau_\nu=1$ surfaces for the three components (red line for the red wing, blue line for the blue wing, and dashed black/white line for the line core).
    \item The colour-collapsed angle averaged radiation field $J_\nu$.
\end{enumerate}
These colour-collapsed plots are very powerful for seeing the spatial collocation of features, but are not colourblind-friendly, thus we also provide versions of the lower two panels with the colour channels exploded as Figure~\ref{Fig:LybColourblind}.
Each channel of these colour-collapsed plots includes Gaussian weighted contributions with a standard deviation of \SI{0.01}{\nano\m}, centred on $\Delta\lambda = 0.025, 0.0, \SI{-0.025}{\nano\m}$ for the red, green, and blue channels, respectively.
For example, a pixel that is predominantly blue has a blue asymmetry, one that is green is dominated by the line core, and one that is magenta contains strong contributions from both line wings, but little from the line core.

Looking at the structure of the atmospheric model, if there are indeed two distinct layers, then we would expect these to be the upper arc above $z=\SI{20}{\mega\m}$ and the lower structure containing the primary condensation below this.
From the contribution function panel, we see that the $\tau_\nu=1$ layer in the line core traces this upper arc other than in the $[0.75, 1.75]\,\si{\mega\m}$ slit position range.
The $\tau_\nu=1$ lines for the blue and red wings move between the upper and lower surfaces much more frequently, but for the majority of the $[-3, 3]\,\si{\mega\m}$ slit position range trace the upper surface of the lower structure.

The contribution function in the primary condensation around slit position \SI{0}{\mega\m} is mostly pink, highlighting the contribution to the line wings from this region, whereas the contribution to the line core along this ray is in the upper arc around $z=\SI{26}{\mega\m}$.
Moving towards the $-x$ side of the primary condensation, the contribution function turns mostly yellow, indicating contributions to the red wing and line core, whilst moving towards the $+x$ side, the contribution function becomes primarily blue, indicating blue wing contributions.

Considering now the strong blue asymmetry in the line profile in the $[-3.5, 0.5]\,\si{\mega\m}$ range, and inspecting the ``arm'' feature on the contribution function panel we see that, with the exception of the blobs around slit position $\SI{-1.75}{\mega\m}$, the $\tau_\nu=1$ surface for the blue wing always lies below the red wing.
The centre of the ``arm'' feature is primarily blue, with a yellow upper edge, and red lower edge.
The blue wing over this range is forming in the core of this structure, whilst increased opacity in the red wing appears to be attenuating deeper contributions.

The bottom panel shows a colour-collapsed representation of $J_\nu$ across the \Lyb{} line.
As the model is illuminated by a line in emission relative to the neighbouring continuum, we see a predominantly green image.
The effects of radiation dilution are clearly apparent when considering the variation in brightness along $z$ at the edge of the image.
Additionally, we note that the structure is casting shadows: these structures will interact radiatively, both pumping and shading each other.
Shadows are also cast by the dark blobs around slit position $\SI{-1.75}{\mega\m}$, and the thicker features of the outer arc.
The cool primary condensation appears dark compared to the surrounding free-space (due to opacity effects blocking incoming radiation), and has a pink tint along the centre of the slit, similarly to the contribution function.
The ``arm'' and lower $+x$ regions have a distinct blue cast, due the the enhancement of the blue wing, the latter of which is at too great an optical depth to contribute significantly to the outgoing spectrum.
The lower $-x$ region has a yellow cast, indicating the enhancement in the red wing here, also at too great an optical depth to strongly influence the outgoing spectrum.

The brightening in the red wing around slit position \SI{-0.75}{\mega\m}, corresponds to a strong enhancement in the red contribution function and $J_\nu$ from layers around $z=\SI{13}{\mega\m}$ and $z=\SI{15}{\mega\m}$ on the edge of the primary condensation.
The deeper layer is notable by being significantly brighter in $J_\nu$ than the rest of the primary condensation.
It corresponds to a region of significantly enhanced pressure in the atmospheric model.

Correlating the previously discussed emission features with the atmospheric model, we see that all emission in the line originates below $\sim\SI{100}{\kilo\kelvin}$, as expected.
The pressure in the primary condensation and ``arm'' are much higher than in the overlying upper arc.
This high pressure, combined with the fluid motions and increased irradiation at this height, leads to the the production of a strong and broad emission line, which is then attenuated by the optically-thick much narrower line formed in the upper arc.
This behaviour is somewhat analogous to the multi-threaded \Lyb{} models of \citet{gunar_lyman-line_2008} where asymmetries were created by stacking multiple prominence threads along the line-of-sight.
\Dex{} enables this kind of study but with self-consistent radiative interaction between these threads.

\subsection{Performance, Memory Consumption, \& Parameters}\label{Sec:DexPerformance}

\begin{table}
\begin{tabular}{lll}
\hline
Code Region                    & Total Time (s)                         & Runtime Fraction      \\ \hline
\multicolumn{1}{l|}{Cascade 5} & \multicolumn{1}{l|}{$4.23\times 10^5$} & 63.6 \%               \\
\multicolumn{1}{l|}{Cascade 4} & \multicolumn{1}{l|}{$1.42\times 10^5$} & 21.4 \%               \\
\multicolumn{1}{l|}{Cascade 3} & \multicolumn{1}{l|}{$5.43\times 10^4$} & 8.2 \%                \\
\multicolumn{1}{l|}{Cascade 2} & \multicolumn{1}{l|}{$1.69\times 10^4$} & 2.5 \%                \\
\multicolumn{1}{l|}{Cascade 1} & \multicolumn{1}{l|}{$5.52\times 10^3$} & 0.8 \%                \\
\multicolumn{1}{l|}{Cascade 0} & \multicolumn{1}{l|}{$3.11\times 10^3$} & 0.5 \%                \\
\multicolumn{1}{l|}{Statistical Equilibrium} & \multicolumn{1}{l|}{$3.99\times 10^2$}    & 0.05 \% \\
\multicolumn{1}{l|}{Charge Conservation}     & \multicolumn{1}{l|}{$5.27\times 10^3$}    & 0.8 \%  \\
\multicolumn{1}{l|}{Pressure Conservation}   & \multicolumn{1}{l|}{$3.98\times 10^{-1}$} & 0 \%    \\ \hline
\end{tabular}
\caption{Timing breakdown of the primary code regions in the \Dex{} solution of the \citet{Jenkins2021} snapshot.}
\label{Tab:Timing}
\end{table}

This model, with a $z\times x$ resolution of $3072 \times 2048$ voxels was solved using an NVIDIA RTX A6000 48\,GB in 185\,GPU hours.
A breakdown of the runtime across the different cascades and population updates is shown in Table~\ref{Tab:Timing}.
Six cascades were used, with a branching factor $\alpha=2$, with a radiance interval length of 1.5\,voxels and 4 angular samples on cascade 0.
For inclination relative to the $y$-axis, a Gauss-Radau quadrature with 8 directions was used.
The bilinear fix was not used as, by inspection, it was not found necessary.
We note that the higher cascades dominate the calculation time, due to the computation cost of the long intervals.
Additionally, the charge conservation step is much slower than expected as it was computed on the CPU rather than GPU here.
Convergence took 378 iterations to a tolerance of $10^{-3}$.

This snapshot was the least sparse of those analysed by \citet{Jenkins2021}; many of the \SI{10}{\gauss} snapshots have $\sim5\,\%$ active cells, and can be solved a few GPU hours.
Solving wavelengths in batches of 4 to minimise GPU warp divergence required 18.5\,GB of GPU memory: reducing the batch size will affect performance but enable this state-of-the-art simulation to be run on a single consumer 8\,GB GPU.
Host (CPU) memory consumption was 17\,GB, primarily due to the storage of the full $J$ array, which was paged on and off the GPU for each wavelength batch.

\section{Concluding Remarks}

We have presented radiance cascades, a novel framework for efficiently computing the radiation field throughout a domain by exploiting the internal structure of this function (through a property termed the penumbra criterion) allowing the reuse of ray segments termed radiance intervals.
Radiance cascades exploit the observation that when considering shells around a point, nearby sources can be resolved with low angular resolution, but vary rapidly in space, while distant sources require high angular resolution, but vary slowly in space.

Additionally, we have shown initial results from \Dex{}, a GPU-accelerated implementation of this technique, performing a state-of-the-art non-LTE synthesis from a $3072\times 2048$ prominence snapshot by \citet{Jenkins2021}, with no evidence of the ray effects that plague short characteristics methods in finely-structured MHD models.
The difference between the \Dex{} and plane-parallel \Pw{} results, especially in lines with high optical thickness such as \Lyb{}, highlights the importance of multidimensional radiative transfer effects in modern finely-structured MHD models, ensuring the consistency between $J$ in the prominence and filament projections.
These are key not only for forward modelling of MHD models but also for future studies considering the radiative losses (and gains) due to optically thick transitions, including correctly accounting for these structures shadowing each other.

Good agreement was found in the overall morphology of the optical spectra (\CaK{} and \Ha{}) between \Dex{} and \Pw{}, however significant variations in intensity were found, particularly from a prominence view.
The \Ha{} proxy of \citet{heinzel_fast_2015} continues to perform admirably for its low computational cost against the two-dimensional radiative transfer results presented here, although not all details present in the \Dex{} model were captured.

The current implementation in \Dex{} is not optimal, and many avenues for improvement exist, including the inclusion of partial frequency redistribution effects through an extension of the Jacobian-Free Newton-Krylov method of \citet{arramy_jacobian-free_2024}, or potentially employing the radiance cascades probe structure as a form of geometric multigrid to accelerate non-LTE convergence \citep[e.g.][]{Stepan2013,Bjorgen2017}.
Ray effects due to short characteristics methods \emph{will} appear in atmospheres with sharp geometric transitions between optically-thin and -thick structures, such as those with overlying arcade structure or around fibrils, although these may be potentially difficult to spot in three-dimensional models, due to visualisation difficulties.
It is also important to correctly include the effective albedo of these structures in models of the lower atmosphere.
We believe that the radiance cascades method provides a key stepping stone towards tractable multidimensional non-LTE radiative transfer and detailed radiative losses in finely structured atmospheric models.

\section*{Acknowledgements}

CMO is grateful to the Royal Astronomical Society's Norman Lockyer fellowship, and the University of Glasgow's Lord Kelvin/Adam Smith Leadership fellowship for financially supporting this work.
They are also grateful to Dr Jack Jenkins for providing the high-resolution snapshot and the encouragement to pursue this project, Dr Ivan Milić for numerical radiative transfer discussions, Dr Tiago Pereira for feedback on the CRTAF draft standard, Prof. Petr Heinzel for guidance on radiative transfer specificities in solar isolated structures, and to the members of the Graphics Programming Discord server's Radiance Cascades thread for insightful discussion on how to implement and improve this technique.

%%%%%%%%%%%%%%%%%%%%%%%%%%%%%%%%%%%%%%%%%%%%%%%%%%
\section*{Data Availability}

All of the data and software underlying this article are publicly available.
The analysis scripts, and inputs and outputs from \Dex{} for the snapshot of \citet{Jenkins2021} are available at \url{https://doi.org/10.5281/zenodo.13375365}.



\bibliographystyle{rasti}
\bibliography{refs}

%%%%%%%%%%%%%%%%%%%%%%%%%%%%%%%%%%%%%%%%%%%%%%%%%%

%%%%%%%%%%%%%%%%% APPENDICES %%%%%%%%%%%%%%%%%%%%%

\appendix

\section{Colour-Collapsed Plot Channels}

\begin{figure*}
	\centering
    \includegraphics[width=\textwidth]{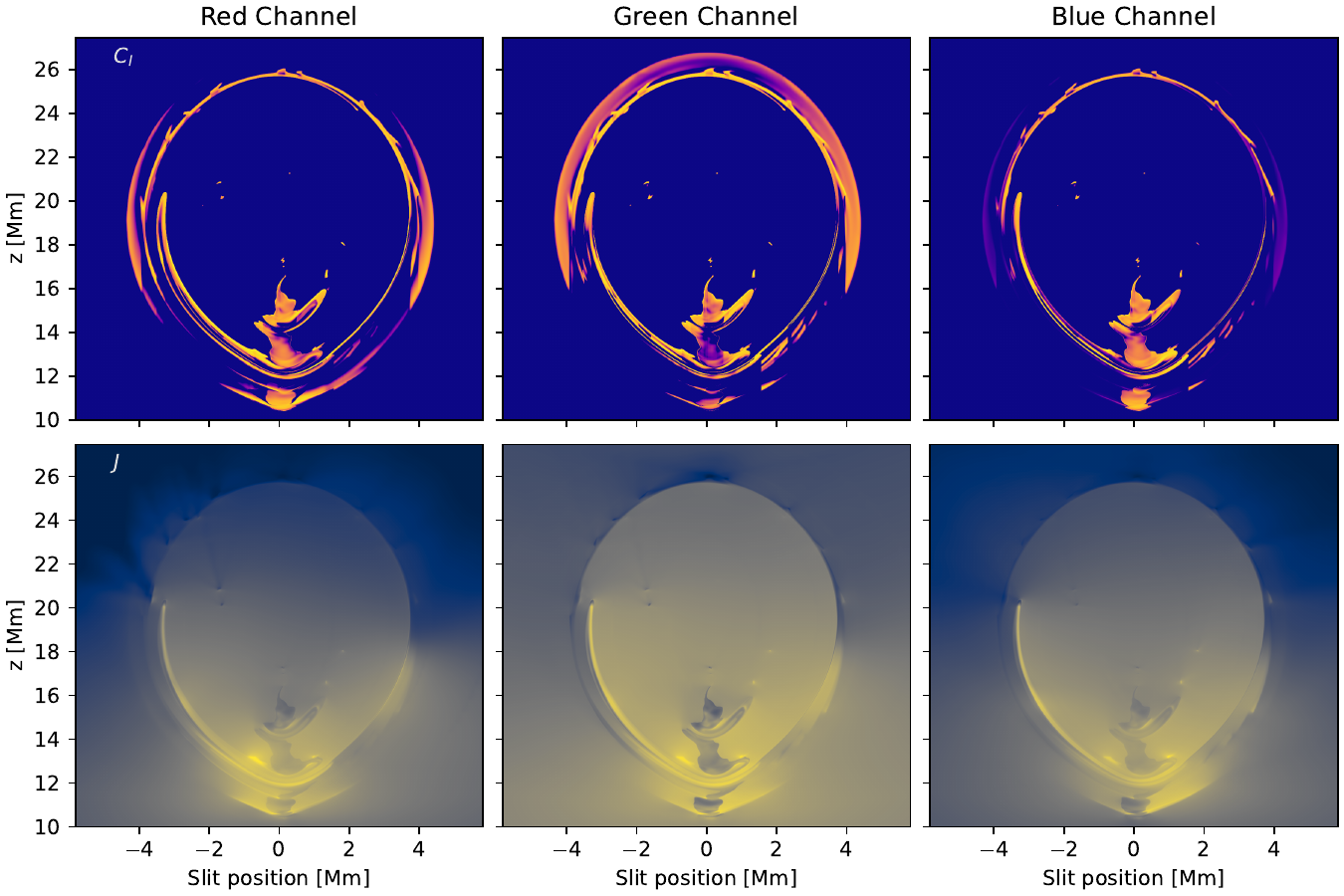}
    \caption{Explodes the colour channels of the COCOPLOTs ($C_I$ and $J$) in Figure~\ref{Fig:LybFormation} for colourblind legibility. Unfortunately, the information is less legible in this form than as a COCOPLOT due to the difficulty of collocating and comparing relative intensity in separate panels.}
    \label{Fig:LybColourblind}
\end{figure*}

Figure~\ref{Fig:LybColourblind} shows the lower two panels of Figure~\ref{Fig:LybFormation} with the colour channels exploded for colourblind legibility.

%%%%%%%%%%%%%%%%%%%%%%%%%%%%%%%%%%%%%%%%%%%%%%%%%%

% Don't change these lines
\bsp	% typesetting comment
\label{lastpage}
\end{document}